\documentclass[11pt]{article} 

\usepackage[margin=1in]{geometry} 
\usepackage[utf8]{inputenc}
\usepackage[T1]{fontenc}
\usepackage{mathpazo} 

\usepackage{authblk}
\usepackage{comment,xspace}
\usepackage[normalem]{ulem} 

\usepackage{booktabs,multicol,multirow,subcaption}

\usepackage{enumitem} 
\usepackage[dvipsnames]{xcolor} 
\definecolor{ptblue}{RGB}{15,76,129} 
\definecolor{ptemerald}{HTML}{009473} 

\usepackage{tikz} 

\usepackage{amsmath,amsfonts,amssymb,amsthm,bbm,mathtools}

\newcommand*{\diff}[1]{\mathop{}\!\mathrm{d}#1} 

\usepackage[square,sort]{natbib} 
\usepackage{hyperref}
\hypersetup{
linktocpage,
colorlinks=true,
citecolor=ptemerald, 
urlcolor=ptblue, 
linkcolor=Plum, 
}
\usepackage{cleveref} 

\theoremstyle{plain}
\newtheorem{theorem}{Theorem}[section]

\theoremstyle{definition}
\newtheorem{definition}[theorem]{Definition}
\newtheorem{example}[theorem]{Example}
\newtheorem{problem}{Open Question}

\theoremstyle{remark}

\makeatletter 
\newcommand{\EF}[1]{\if\relax\detokenize\expandafter{\@firstofone#1{}}\relax EF\xspace\else EF#1\fi}
\makeatother
\newcommand{\EFOne}{\EF{1}\xspace}
\newcommand{\EFX}{\EF{X}\xspace}
\newcommand{\EFXzero}{\EF{X}$_0$\xspace}
\newcommand{\EFM}{\EF{M}\xspace}
\newcommand{\EFMzero}{\EF{M}$_0$\xspace}
\newcommand{\EFoneM}{\EF{1M}\xspace} 
\newcommand{\EFXM}{\EF{XM}\xspace} 
\newcommand{\MMS}{\textup{MMS}\xspace}
\newcommand{\RW}{Robertson-Webb\xspace}

\newcommand{\length}[1]{\texttt{len}(#1)}

\newcommand{\alloc}{\mathcal{A}}
\newcommand{\bundle}{A}
\newcommand{\resource}{R}
\newcommand{\divisibleItems}{D} 
\newcommand{\indivisibleGoods}{G}
\newcommand{\indivisibleChores}{C}
\newcommand{\indivisibleItems}{O}

\title{Mixed Fair Division: A Survey}
\author[1]{Shengxin Liu}
\author[2]{Xinhang Lu}
\author[2]{Mashbat Suzuki}
\author[2]{Toby Walsh}
\affil[1]{Harbin Institute of Technology, Shenzhen}
\affil[2]{UNSW Sydney}
\affil[ ]{\nolinkurl{sxliu@hit.edu.cn}, \nolinkurl{{xinhang.lu, mashbat.suzuki, t.walsh}@unsw.edu.au}}
\date{}

\begin{document}
\maketitle

\begin{abstract}
Fair division considers the allocation of scarce resources among agents in such a way that every agent gets a fair share.
It is a fundamental problem in society and has received significant attention and rapid developments from the game theory and artificial intelligence communities in recent years.
The majority of the fair division literature can be divided along at least two orthogonal directions: goods versus chores, and divisible versus indivisible resources.
In this survey, besides describing the state of the art, we outline a number of interesting open questions and future directions in three \emph{mixed} fair division settings: (i) indivisible goods and chores, (ii) divisible and indivisible goods (mixed goods), and (iii) indivisible goods with subsidy which can be viewed like a divisible good.
\end{abstract}

\section{Introduction}
\label{sec:intro}

In fair division, we look to allocate resources fairly among agents with possibly heterogeneous preferences over the resources.
Fair division is a fundamental research topic in computational social choice~\citep{BrandtCoEn16,Endriss17,Rothe24}.
It has a long and rich history dating back to the work of \citet{Steinhaus48}, and has attracted ongoing interest from mathematicians, economists, and computer scientists in the past several decades~\citep{AmanatidisAzBi23,Aziz20,BramsTa96,Moulin19,NguyenRo23,RobertsonWe98,Suksompong21,Suksompong25,Walsh20}.
Moreover, fair division methods have been deployed in practice~\citep{BudishCaKe17} and made publicly available~\citep{GoldmanPr15,IgarashiYo23,HanSu24,Shah17}; see also the Adjusted Winner website\footnote{\url{https://pages.nyu.edu/adjustedwinner}} and a Rent Division Calculator\footnote{\url{https://www.nytimes.com/interactive/2014/science/rent-division-calculator.html}}.

The vast majority of fair division literature can be divided along two orthogonal directions according to:
\begin{itemize}
\item the (in)divisibility of the resources, and
\item agents' valuations over the resources.
\end{itemize}
Specifically, in the former case, the resource is either \emph{divisible} or \emph{indivisible}, and in the latter case, the resource consists of either \emph{goods} (positively valued) or \emph{chores} (negatively valued).
In many real-world scenarios, however, the resources to be allocated may be a mixture of different types.
Our first example demonstrates a mixture of (indivisible) \emph{goods and chores}: when distributing household tasks, some family member may enjoy cooking while others may find it torturous.
The next example touches on a mixture of \emph{divisible and indivisible} goods: when dividing up an estate or assets in a divorce, we usually have divisible goods like money, as well as indivisible goods like houses, cars, paintings, etc.
It may also be that monetary compensation (a.k.a.\ subsidies) could help circumvent unfair allocations of indivisible inheritances.
Classic fairness notions or algorithmic methods that work well with a single type of resources may not fare well in the aforementioned scenarios concerning mixed types of resources.

In this survey, we discuss fair division with mixed types of resources, which has received growing attention in recent years, and focus on three mixed fair division domains:
\begin{itemize}
\item \Cref{sec:indivisible-items} considers fair division of indivisible goods and chores, in which each agent may have positive, zero, or negative valuation over each item;

\item \Cref{sec:mixed-div-ind} focuses on fair division of mixed divisible and indivisible goods (\emph{mixed goods});

\item \Cref{sec:subsidy} focuses on fair division of indivisible goods with subsidy.
\end{itemize}

Clearly, the first and second domains relax one of the two orthogonal directions mentioned earlier.
The second and third domains share some similarity in the sense that subsidy could be viewed as a divisible good; the key difference lies in how they approach fairness.
In \Cref{sec:mixed-div-ind}, both the divisible and indivisible goods are \emph{fixed} in advance and we find \emph{approximately} fair allocations.
In \Cref{sec:subsidy}, we allocate indivisible goods but introduce some \emph{additional} amount of money in order to satisfy \emph{exact} fairness.

This survey outlines new fairness notions and related theoretical results that are addressed in the above mixed fair division settings as well as highlights a number of major open questions and interesting directions for future research.

\section{Preliminaries}
\label{sec:prelim}

For each $k \in \mathbb{N}$, let $[k] \coloneqq \{1, 2, \dots, k\}$.
Denote by~$N = [n]$ the set of~$n$ agents to whom we allocate some resource~$\resource$, which may, e.g., consist of indivisible goods and chores (\Cref{sec:prelim:indivisible-items}) or be a mix of divisible and indivisible goods (\Cref{sec:prelim:mixed-div-ind}).
An allocation $\alloc = (\bundle_1, \bundle_2, \dots, \bundle_n)$ assigns bundle~$\bundle_i$ to agent~$i \in N$ and $\bundle_i \cap \bundle_j = \emptyset$ for all $i \neq j$; note that~$\bundle_i$ can be empty.
An allocation is said to be \emph{complete} if the entire resource is allocated, i.e., $\bigcup_{i \in N} \bundle_i = R$, and \emph{partial} otherwise.
Unless specified otherwise, we assume allocations considered in this survey are complete.

\subsection{Cake Cutting}
\label{sec:prelim:cake}

When resource~$\resource$ is heterogeneous and infinitely \emph{divisible}, the corresponding problem is commonly known as \emph{cake cutting}~\citep{BramsTa96,LindnerRo24,Procaccia16,RobertsonWe98}.
We will use the terms ``cake'' and ``divisible goods'' interchangeably.
The cake, denoted by~$\divisibleItems$, is represented by the normalized interval~$[0, 1]$.
A \emph{piece of cake} is a union of finitely many disjoint (closed) intervals.
Each agent~$i \in N$ is endowed with an integrable \emph{density function} $f_i \colon [0, 1] \to \mathbb{R}_{\geq 0}$, capturing how the agent values each part of the cake.
Given a piece of cake~$S \subseteq [0, 1]$, agent~$i$'s utility over~$S$ is defined as $u_i(S) \coloneqq \int_{S} f_i(x) \diff{x}$.
Denote by $(\divisibleItems_1, \divisibleItems_2, \dots, \divisibleItems_n)$ the allocation of cake~$\divisibleItems$.
In order to access agents' density functions, the cake-cutting literature usually adopts the \emph{\RW (RW) query model}~\citep{RobertsonWe98}, which allows an algorithm to interact with the agents via the following two types of queries:
\begin{itemize}
\item $\textsc{Eval}_i(x, y)$ returns~$u_i([x, y])$;
\item $\textsc{Cut}_i(x, \alpha)$ asks agent~$i$ to return the leftmost point~$y$ such that~$u_i([x, y]) = \alpha$, or state that no such~$y$ exists.
\end{itemize}

\paragraph{Homogeneous Cake}

A \emph{homogeneous} cake is a special case in which each density function~$f_i$ takes on some constant value.
Put differently, every agent values all pieces of equal length the same.
\emph{Money}, for example, can be viewed as a homogeneous cake that is valued the same by all agents.

\subsection{Mixed Indivisible Goods and Chores}
\label{sec:prelim:indivisible-items}

Discrete fair division, in which resource~$R$ consists of \emph{indivisible} items, has received considerable attention in the last two decades, especially for allocating \emph{goods}; see, e.g., \citep{AmanatidisAzBi23,Moulin19,NguyenRo23,Suksompong21,Suksompong25} for an overview of the most recent developments.

We present here a general model where an agent may have a positive, zero, or negative utility for each indivisible item.
Specifically, denote by~$\indivisibleItems = [m]$ the set of~$m$ indivisible items.
An (indivisible) \emph{bundle} is a subset of~$\indivisibleItems$.
Each agent~$i \in N$ is endowed with a \emph{utility function} $u_i \colon 2^\indivisibleItems \to \mathbb{R}$ such that $u_i(\emptyset) = 0$, capturing how the agent values each bundle of the items.
For an item~$o \in \indivisibleItems$, we will write~$u_i(o)$ instead of~$u_i(\{o\})$ for simplicity.
A utility function~$u$ is said to be \emph{additive} if $u(O') = \sum_{o \in O'} u(o)$ for any~$O' \subseteq \indivisibleItems$.
Unless specified otherwise, we assume agents have additive utilities in this survey.
Let $\mathcal{\indivisibleItems} = (\indivisibleItems_1, \indivisibleItems_2, \dots, \indivisibleItems_n)$ denote the allocation of items~$\indivisibleItems$.

We say that an item~$o \in \indivisibleItems$ is a \emph{good} (resp., \emph{chore}) for agent~$i$ if $u_i(o) \geq 0$ (resp., $u_i(o) \leq 0$), and let $\indivisibleGoods_i$ (resp., $\indivisibleChores_i$) be the set of goods (resp., chores) for agent~$i$.
In other words, for each item, agents have \emph{subjective} opinions on whether the item is a good or a chore.
An item is said to be an \emph{objective good} (resp., \emph{objective chore}) if the item is a good (resp., chore) for all agents.
The presented model includes scenarios where \emph{all} items are objective goods (resp., objective chores), which we will specifically refer to as an indivisible-goods (resp., indivisible-chores) setting.

\paragraph{(Doubly-)Monotonic Utilities}
While we mostly focus on additive utilities, we will identify some results that still hold with a larger class of utility functions.
The utility function~$u_i$ of agent~$i \in N$ is said to be \emph{doubly-monotonic} if agent~$i$ can partition the items as $\indivisibleItems = \indivisibleGoods_i \sqcup \indivisibleChores_i$ such that for any item~$o \in \indivisibleItems$ and for any bundle~$O' \subseteq \indivisibleItems \setminus \{o\}$,
\begin{itemize}
\item $u_i(O' \cup \{o\}) \geq u_i(O')$ if $o \in \indivisibleGoods_i$, and
\item $u_i(O' \cup \{o\}) \leq u_i(O')$ if $o \in \indivisibleChores_i$.
\end{itemize}

In the indivisible-goods (resp., indivisible-chores) setting, all agents~$i \in N$ have \emph{monotonically non-decreasing (resp., non-increasing)} utility functions, that is, $u_i(S) \leq u_i(T)$ (resp., $u_i(S) \geq u_i(T)$) for any bundles $S \subseteq T \subseteq \indivisibleItems$.

\subsection{Mixed Divisible and Indivisible Goods}
\label{sec:prelim:mixed-div-ind}

We now introduce a fair division model with both divisible \emph{and} indivisible goods (henceforth \emph{mixed goods} for short).
In the mixed-goods setting, resource~$\resource$ consists of a cake~$D = [0, 1]$ and a set of indivisible \emph{goods}~$\indivisibleItems = [m]$.
Each agent~$i \in N$ has a density function~$f_i$ over the cake as defined in \Cref{sec:prelim:cake} and an additive utility function~$u_i$ over indivisible goods~$\indivisibleItems$.
Denote by $\alloc = (\bundle_1, \bundle_2, \dots, \bundle_n)$ the allocation of mixed goods, where $\bundle_i  = \divisibleItems_i \cup \indivisibleItems_i$ is the \emph{bundle} allocated to agent~$i$.
Agent~$i$'s utility is defined as $u_i(A_i) \coloneqq u_i(\divisibleItems_i) + u_i(\indivisibleItems_i)$.
Further discussion about the model, including the definitions of fairness notions and other extensions, is provided in \Cref{sec:mixed-div-ind}.

\section{Solution Concepts}
\label{sec:fairness-notions}

Before introducing fairness concepts considered in this survey, we first define Pareto optimality, an economic efficiency notion that is fundamental in the context of fair division.

\begin{definition}[PO]
Given an allocation $\alloc = (\bundle_i)_{i \in N}$, another allocation $\alloc' = (\bundle'_i)_{i \in N}$ is said to be a \emph{Pareto improvement} if $u_i(\bundle'_i) \geq u_i(\bundle_i)$ for all~$i \in N$ and $u_j(\bundle'_j) > u_j(\bundle_j)$ for some~$j \in N$.
Alternatively, we say that~$\alloc$ is \emph{Pareto dominated} by~$\alloc'$.
An allocation is said to satisfy \emph{Pareto optimality (PO)} if it does not admit a Pareto improvement.
\end{definition}

In what follows, we first introduce comparison-based fairness notions (i.e., envy-freeness relaxations) in \Cref{sec:fairness-notions:EF}, followed by fair-share-based notions (e.g., proportionality and maximin share guarantee) in \Cref{sec:fairness-notions:PROP,sec:fairness-notions:MMS}.

\subsection{(Approximate) Envy-Freeness}
\label{sec:fairness-notions:EF}

\emph{Envy-freeness}---the epitome of fairness, as \citet{Procaccia20} put it---requires that every agent likes her own bundle at least as much as the bundle given to any other agent.

\begin{definition}[\EF{}~\citep{Tinbergen30,Foley67,Varian74}\protect\footnote{We refer the interested readers to the paper of \citet{HeilmannWi21} for more discussion on the work of \citet{Tinbergen30}.}]
An allocation $(\bundle_1, \bundle_2, \dots, \bundle_n)$ is said to satisfy \emph{envy-freeness (\EF{})} if for any pair of agents~$i, j \in N$, $u_i(A_i) \geq u_i(A_j)$.
\end{definition}

In cake cutting, an envy-free cake division always exists~\citep{Su99}.
This can also be seen from a result of \citet{Alon87}.
A $k$-partition $(\divisibleItems_1, \divisibleItems_2, \dots, \divisibleItems_k)$ of cake~$\divisibleItems$ is said to be \emph{perfect} if each agent~$i \in N$ values all pieces equally, that is, $u_i(\divisibleItems_j) = \frac{u_i(\divisibleItems)}{k}$ for all~$i \in N$ and~$j \in [k]$.
\citet{Alon87} showed that a perfect partition of the cake always exists for any set of agents and any~$k \in \mathbb{N}$.
It implies that an envy-free cake division always exists.

An envy-free allocation need not exist when allocating indivisible items.
To circumvent this issue, relaxations of envy-freeness have been proposed and studied.

\begin{definition}[\EFOne~\citep{LiptonMaMo04,Budish11,AzizCaIg22}]
An allocation $(\indivisibleItems_1, \dots, \indivisibleItems_n)$ of indivisible items~$\indivisibleItems$ is said to satisfy \emph{envy-freeness up to one item (\EFOne)} if for every pair of agents $i, j \in N$, either
\begin{itemize}
\item there exists~$\indivisibleItems' \subseteq \indivisibleItems_j$ with $|\indivisibleItems'| \leq 1$ such that $u_i(\indivisibleItems_i) \geq u_i(\indivisibleItems_j \setminus \indivisibleItems')$, or
\item there exists~$\indivisibleItems' \subseteq \indivisibleItems_i$ with $|\indivisibleItems'| \leq 1$ such that $u_i(\indivisibleItems_i \setminus \indivisibleItems') \geq u_i(\indivisibleItems_j)$.
\end{itemize}
\end{definition}

Intuitively, \EFOne requires that when agent~$i$ envies agent~$j$, the envy can be eliminated by either removing some good (in agent~$i$'s view) from agent~$j$'s bundle or removing some chore (again, in agent~$i$'s view) from agent~$i$'s own bundle.
We will introduce a stronger notion than \EFOne.
Before that, we first restrict ourselves to the indivisible-goods setting and strengthen \EFOne in the following sense: any envy should be eliminated even if we remove the least (positively) valued good from the envied bundle.

\begin{definition}[\EFXzero and \EFX for indivisible goods\protect\footnote{\label{ft:EFX0-EFX-nomenclature}The nomenclature of \EFXzero and \EFX is adopted from \citet{KyropoulouSuVo20}.}~\citep{CaragiannisKuMo19,PlautRo20}]
\label{def:EFX_0-EFX-goods}
An indivisible-goods allocation $(\indivisibleItems_1, \indivisibleItems_2, \dots, \indivisibleItems_n)$ is said to satisfy
\begin{itemize}
\item \emph{envy-freeness up to any good (\EFXzero)} if for any pair of agents~$i, j \in N$ and any good~$g \in \indivisibleItems_j$, $u_i(\indivisibleItems_i) \geq u_i(\indivisibleItems_j \setminus \{g\})$;

\item \emph{envy-freeness up to any \emph{positively valued} good (\EFX)} if for any pair of agents~$i, j \in N$ and any good~$g \in \indivisibleItems_j$ such that $u_i(g) > 0$, we have $u_i(\indivisibleItems_i) \geq u_i(\indivisibleItems_j \setminus \{g\})$.
\end{itemize}
\end{definition}

\EFXzero is a stronger variant than \EFX, which in turn imposes a stronger requirement than \EFOne.
For indivisible goods, an \EFXzero (and hence \EFX) allocation always exists for at most three agents \citep{AkramiAlCh23,ChaudhuryGaMe24,PlautRo20}, but the existence of \EFX allocations remains open for four or more agents.
\EFXzero, however, does not fare well with PO~\citep{PlautRo20}.
We will also see such nuances and conflicts in \Cref{sec:mixed-div-ind} when introducing fairness notions in the mixed-goods setting.
With mixed indivisible goods and chores, we define \EFX as follows:

\begin{definition}[\EFX and \EFXzero for indivisible goods and chores~\citep{AzizCaIg22,AzizRe20,HosseiniSiVa23}]
An allocation $(\indivisibleItems_1, \indivisibleItems_2, \dots, \indivisibleItems_n)$ of indivisible goods and chores is said to satisfy
\begin{itemize}
\item \emph{envy-freeness up to any item (\EFXzero)} if for any pair of agents~$i, j \in N$:
\begin{itemize}
\item $u_i(\indivisibleItems_i) \geq u_i(\indivisibleItems_j \setminus \{o\})$ for any~$o \in \indivisibleGoods_i \cap \indivisibleItems_j$, and
\item $u_i(\indivisibleItems_i \setminus \{o\}) \geq u_i(O_j)$ for any~$o \in \indivisibleChores_i \cap \indivisibleItems_i$;
\end{itemize}

\item \emph{envy-freeness up to any non-zero valued item (\EFX)} if for any pair of agents~$i, j \in N$:
\begin{itemize}
\item $u_i(\indivisibleItems_i) \geq u_i(\indivisibleItems_j \setminus \{o\})$ for any~$o \in \indivisibleGoods_i \cap \indivisibleItems_j$ with~$u_i(o) \neq 0$, and
\item $u_i(\indivisibleItems_i \setminus \{o\}) \geq u_i(O_j)$ for any~$o \in \indivisibleChores_i \cap \indivisibleItems_i$ with~$u_i(o) \neq 0$.
\end{itemize}
\end{itemize}
\end{definition}

The envy relations between the agents in an allocation are commonly captured by the \emph{envy graph}, in which the vertices correspond to the agents and there is a directed edge from one agent to another if the former agent envies the latter~\citep{LiptonMaMo04}.
Variants of the envy graph and additional techniques are introduced in many other papers~\citep[e.g.,][]{HalpernSh19,BeiLiLi21,BhaskarSrVa21,AmanatidisAzBi23}.

The following example demonstrates \EFOne, \EFX, and \EFXzero allocations.

\begin{example}
\label{example:prelim}
Consider an example with three agents and four items~$\{o_1, o_2, o_3, o_4\}$.
Agents' valuations are listed below:

\begin{center}
\begin{tabular}{@{}c|*{4}{r}@{}}
\toprule
& $o_1$ & $o_2$ & $o_3$ & $o_4$ \\
\midrule
$u_1$ & $-1$ & $-1$ & $-2$ & $-2$ \\
$u_2$ & $1$ & $1$ & $2$ & $2$ \\
$u_3$ & $1$ & $0$ & $2$ & $1$ \\
\bottomrule
\end{tabular}
\label{tab:prelim-example}
\end{center}

Let us consider the following three allocations:
\begin{center}
\begin{tabular}{@{}l|*{3}{l}@{}}
\toprule
& Agent~$1$ & Agent~$2$ & Agent~$3$ \\
\midrule
Allocation~$\alloc$ & $\{o_2, o_3\}$ & $\{o_1\}$ & $\{o_4\}$ \\
Allocation~$\alloc'$ & $\{o_1\}$ & $\{o_2, o_3\}$ & $\{o_4\}$ \\
Allocation~$\alloc''$ & $\{o_1\}$ & $\{o_3\}$ & $\{o_2, o_4\}$ \\
\bottomrule
\end{tabular}
\end{center}

Allocation~$\alloc$ is \EFOne.
It is not \EFX, because, e.g., $a_2$ still envies $a_1$ when removing $a_2$'s least preferred good from~$\bundle_1$, i.e., $u_2(\bundle_2) = 1 < 2 = u_2(\bundle_1 \setminus \{o_2\})$.

Allocation~$\alloc'$ is \EFX (and thus \EFOne).
In particular, $a_3$'s envy towards $a_2$ can be eliminated by removing $a_3$'s least positively valued good~$o_3$ from $\bundle'_2$, i.e., $u_3(\bundle'_3) = 1 \geq u_3(\bundle'_2 \setminus \{o_3\}) = 0$.
It is not \EFXzero because $o_2$ is $a_3$'s least valued good in $\bundle'_2$ but $u_3(\bundle'_3) = 1 < 2 = u_3(\bundle'_2 \setminus \{o_2\})$.

Allocation~$\alloc''$ is \EFXzero (and hence \EFX and \EFOne).
This can be seen from the fact that
\begin{itemize}
\item $a_1$ does not envy~$a_2$ or~$a_3$, nor is envied by any agent;
\item $a_2$'s envy towards~$a_3$ can be eliminated by removing~$a_2$'s least valuable good~$o_2$ from~$\bundle''_3$;
\item $a_3$'s envy towards~$a_2$ can be eliminated by removing~$a_3$'s least valuable good~$o_3$ from~$\bundle''_2$.
\end{itemize}
\end{example}

We defer our discussion on relaxations of envy-freeness in the mixed-goods model to \Cref{sec:mixed-div-ind:EFM}.
It is worth noting that \citet{BeiLiLi21} proposed a notion that naturally combines envy-freeness and \EFOne together and is guaranteed to be satisfiable.

\subsection{Proportionality}
\label{sec:fairness-notions:PROP}

We now introduce fair-share-based notions.
Our first fairness notion is \emph{proportionality}, which requires that each agent receives value at least~$1/n$ of her value for the entire set of resource~$\resource$.
For additive utilities, proportionality is weaker than envy-freeness.

\begin{definition}[PROP~\citep{Steinhaus48}]
An allocation~$\alloc = (\bundle_i)_{i \in N}$ is said to satisfy \emph{proportionality (PROP)} if for every agent~$i \in N$, $u_i(\bundle_i) \geq \frac{u_i(\resource)}{n}$.
\end{definition}

A proportional cake division always exists~\citep{Steinhaus48}.
This is not the case when allocating indivisible items.
As a result, relaxations of proportionality have been studied.
For instance, PROP1 defined below requires that each agent receives her proportional share by either obtaining an additional good (from other agents' bundles) or removing some chore from her own bundle.

\begin{definition}[PROP1 and PROPX~\citep{AzizCaIg22,AzizMoSa20,ConitzerFrSh17,Moulin19}]
An allocation~$(\indivisibleItems_i)_{i \in N}$ of indivisible goods and chores~$\indivisibleItems$ is said to satisfy
\begin{itemize}
\item \emph{proportionality up to one item (PROP1)} if for each agent~$i \in N$,
\begin{itemize}
\item $u_i(\indivisibleItems_i) \geq \frac{u_i(\indivisibleItems)}{n}$,
\item $u_i(\indivisibleItems_i \cup \{o\}) \geq \frac{u_i(\indivisibleItems)}{n}$ for some~$o \in \indivisibleItems \setminus \indivisibleItems_i$, or
\item $u_i(\indivisibleItems_i \setminus \{o\}) \geq \frac{u_i(\indivisibleItems)}{n}$ for some~$o \in \indivisibleItems_i$;
\end{itemize}

\item \emph{proportionality up to any item (PROPX)} if for each agent~$i \in N$,
\begin{itemize}
\item $u_i(\indivisibleItems_i \setminus \{o\}) \geq \frac{u_i(\indivisibleItems)}{n}$ for all~$o \in \indivisibleItems_i$ with $u_i(o) < 0$, and
\item $u_i(\indivisibleItems_i \cup \{o\}) \geq \frac{u_i(\indivisibleItems)}{n}$ for all~$o \in \indivisibleItems \setminus \indivisibleItems_i$ with $u_i(o) > 0$.
\end{itemize}
\end{itemize}
\end{definition}

It follows from the definitions that PROP $\implies$ PROPX $\implies$ PROP1.
With mixed indivisible goods and chores, \EFOne implies PROP1~\citep{AzizCaIg22}.
With only indivisible goods, \EFX and PROPX are not comparable to each other.
First, it can be seen from the following example that PROPX does not imply \EFX.
Consider two agents, two indivisible goods, and both agents value each good at~$1$.
Allocating all goods to a single agent satisfies PROPX.
The allocation, however, is not \EFX because the empty-handed agent envies the other agent even if any good is removed from the latter agent's bundle.
Next, \EFX does not imply PROPX either.
This can be seen from the fact that an \EFX allocation of indivisible goods always exists for three agents~\citep{ChaudhuryGaMe24}, but there exist three-agent instances in which PROPX allocations do not exist~\citep{Moulin19,AzizMoSa20}.
On the contrary, with only indivisible chores, \EFX implies PROPX~\citep{AzizLiMo24}.
Moreover, unlike the indivisible-goods setting, a PROPX allocation of indivisible chores always exist and can be computed efficiently~\citep{Moulin19,LiLiWu22,AzizLiMo24}.

Below, we demonstrate PROP, PROPX and PROP1 allocations.

\begin{example}
Consider the instance in \Cref{example:prelim}.
The proportional share of agent~$1$ (respectively, $2$ and~$3$) is~$-2$ (respectively, $2$ and~$4/3$).
Allocation $(\emptyset, \{o_1, o_2, o_4\}, \{o_3\})$ is proportional.
Allocation $(\{o_2, o_3\}, \{o_1\}, \{o_4\})$ is PROPX but not proportional:
\begin{itemize}
\item When removing agent~$1$'s most-valued chore~$o_2$ from her bundle, she reaches her proportional share of~$-2$.
\item When adding agent~$2$'s least-valued good~$o_2 \notin \bundle_2$ to her bundle, she reaches her proportional share of~$2$.
\item When adding agent~$3$'s least-valued and positively-valued good~$o_1 \notin \bundle_1$, she reaches her proportional share of~$4/3$.
\end{itemize}
Allocation~$(\{o_3, o_4\}, \{o_1\}, \{o_2\})$ is PROP1 but not PROPX.
Note that agent~$3$'s bundle does not meet the PROPX criterion.
\end{example}

\subsection{Maximin Share Guarantee}
\label{sec:fairness-notions:MMS}

Finally, we introduce another well-known fair-share-based notion called the maximin share (MMS) guarantee, and present below a unified definition working for mixed fair division settings.
The MMS guarantee is inspired by generalizing the \emph{divide-and-choose} procedure which produces an (almost) envy-free allocation with two agents~\citep[see, e.g.,][]{Budish11}.

\begin{definition}[$\alpha$-MMS~\citep{Budish11,KulkarniMeTa21,BeiLiLu21}]
\label{def:MMS}
Let~$\Pi_n(\resource)$ be the set of $n$-partitions of resource~$\resource$.
The \emph{maximin share (MMS)} of agent~$i$ is defined as
\[
\MMS_i(n, \resource) = \max_{(P_1, \dots, P_n) \in \Pi_n(\resource)} \min_{j \in [n]} u_i(P_j).
\]
Any partition for which this maximum is attained is called an \emph{MMS partition} of agent~$i$.
We will simply refer to $\MMS_i(n, \resource)$ as~$\MMS_i$ when the context of parameters is clear.

An allocation $\alloc = (\bundle_1, \bundle_2, \dots, \bundle_n)$ of resource~$\resource$ is said to satisfy the \emph{$\alpha$-approximate maximin share guarantee ($\alpha$-MMS)}, for some~$\alpha \in (0, 1]$, if for every~$i \in N$,
\[
u_i(\bundle_i) \geq \min \left\{ \alpha \cdot \MMS_i(n, \resource), \frac{1}{\alpha} \cdot \MMS_i(n, \resource) \right\}.
\]
That is, $\alpha$-MMS requires that $u_i(\bundle_i) \geq \alpha \cdot \MMS_i(n, \resource)$ when agent~$i$ has a non-negative maximin share (i.e., $\MMS_i(n, \resource) \geq 0$) and $u_i(\bundle_i) \geq \frac{1}{\alpha} \cdot \MMS_i(n, \resource)$ when the agent has a negative maximin share (i.e., $\MMS_i(n, \resource) < 0$).
When $\alpha = 1$, we simply refer to $1$-MMS as the MMS guarantee.
\end{definition}

We use the following example to demonstrate agents' MMS values (and their corresponding MMS partitions), as well as approximate-MMS allocations.

\begin{example}
Consider the instance in \Cref{example:prelim}.
Below, we list each agent's maximin share and their corresponding MMS partition:
\begin{itemize}
\item $\MMS_1 = -2$, and $(\{o_1, o_2\}, \{o_3\}, \{o_4\})$ is the MMS partition of agent~$1$;
\item $\MMS_2 = 2$, and $(\{o_1, o_2\}, \{o_3\}, \{o_4\})$ is the MMS partition of agent~$2$;
\item $\MMS_3 = 1$, and $(\{o_1\}, \{o_2, o_3\}, \{o_4\})$ is an MMS partition of agent~$3$.
\end{itemize}

Consider allocations~$\alloc$, $\alloc'$ and~$\alloc''$ specified in \Cref{example:prelim}.
Allocation~$\alloc$ is $\frac{1}{2}$-MMS but not $(\frac{1}{2} + \varepsilon)$-MMS for any~$\varepsilon > 0$, because each agent gets a utility of at least one half of their own MMS value and agent~$2$ gets a utility of exactly one half of her MMS value:
\begin{itemize}
\item $u_1(\{o_2, o_3\}) = -3 \geq -4 = \min \left\{ \frac{1}{2} \cdot (-2), \frac{1}{1/2} \cdot (-2) \right\}$;
\item $u_2(\{o_1\}) = 1 = \min \left\{ \frac{1}{2} \cdot 2, \frac{1}{1/2} \cdot 2 \right\}$;
\item $u_3(\{o_4\}) = 1 \geq \frac{1}{2} = \min \left\{ \frac{1}{2} \cdot 1, \frac{1}{1/2} \cdot 1 \right\}$.
\end{itemize}
Similarly, it can be verified that both allocations~$\alloc'$ and~$\alloc''$ satisfy the MMS guarantee.
\end{example}

If an $\alpha$-MMS allocation is guaranteed to exist, an $\alpha$-MMS and PO allocation always exists as well, because an $\alpha$-MMS allocation which does not admit a Pareto improvement is PO.
In fact, for a fair-share-based notion, a Pareto improvement preserves the fairness notion.
Note, however, that it is co-NP-complete to decide whether a given allocation is PO~\citep{deKeijzerBoKl09,AzizBiLa19}.

As we have seen in \Cref{def:MMS}, the (approximate) MMS guarantee can be naturally defined for settings involving indivisible goods and chores by letting $\resource = \indivisibleItems$ (\Cref{sec:prelim:indivisible-items}) or mixed goods by letting $\resource = \divisibleItems \cup \indivisibleItems$ (\Cref{sec:prelim:mixed-div-ind}).
We will discuss in \Cref{sec:indivisible-items:MMS,sec:mixed-div-ind:MMS} the recent results on approximate MMS guarantee in respective settings.

\section{Mixed Indivisible Goods and Chores}
\label{sec:indivisible-items}

This section is concerned with the fair division of mixed indivisible goods and chores described in \Cref{sec:prelim:indivisible-items}.
We will discuss approximate envy-free allocations in \Cref{sec:indivisible-items:EF}, followed by discussions of MMS in \Cref{sec:indivisible-items:MMS}.

\subsection{Envy-freeness Relaxations}
\label{sec:indivisible-items:EF}

Chores might be viewed simply as ``negative'' goods.
Ordinal methods for allocating goods can then be used directly by simply ordering chores after goods.
However, certain properties are lost in such an approach.
The fundamental problem is an asymmetry between goods and chores: an absence of goods is the worst possible outcome, but an absence of chores is the best possible outcome.

We observe this (breakdown in) duality, for example, when allocating goods in a round-robin fashion.
The \emph{round-robin} algorithm works by arranging the agents in an arbitrary order, and letting each agent in the order choose her favourite good from the remaining goods.
With additive utilities, this is guaranteed to return an \EFOne allocation~\citep{CaragiannisKuMo19}.
The proof is simple.
If Alice picks before Bob, then Alice can always pick a more valuable item to her than Bob next picks.
But if Alice picks after Bob, we ignore the first item that Bob picks, and now the item that Alice picks is always more valuable to Alice than the next item picked by Bob.
This argument breaks when we have \emph{both goods and chores}, and the allocation returned may not be \EFOne.

\begin{example}[The round-robin algorithm does not satisfy \EFOne~\citep{AzizCaIg22}]
\label{ex:rr-is-not-EF1}
Consider the following instance with two agents who have identical utilities over four items:
\begin{center}
\begin{tabular}{@{}l*{4}{c}@{}}
\toprule
& $o_1$ & $o_2$ & $o_3$ & $o_4$ \\
\midrule
Alice, Bob: & $2$ & $-3$ & $-3$ & $-3$ \\
\bottomrule
\end{tabular}
\end{center}
Assume without loss of generality that Alice chooses first and Bob next.
Then, Alice gets the positively valued good~$o_1$ and one chore (say,~$o_3$), whereas Bob gets the other two chores.
As a result, Bob remains envious even if one item is removed from the bundles of Alice and Bob.
\end{example}

We can, however, modify the round-robin algorithm to ensure the allocation returned is \EFOne for mixed indivisible goods and chores.
At a high level, the \emph{double round-robin algorithm} of \citet{AzizCaIg22} applies the round-robin algorithm twice as follows: Agents first pick \emph{objective} chores in a round-robin fashion; we then \emph{reverse} the picking order of the agents for the remaining items and let the agents take turns to pick their favourite \emph{good}.
We demonstrate the algorithm by applying it to \Cref{ex:rr-is-not-EF1}.
First, we introduce one dummy chore~$o$ where both Alice and Bob value~$o$ at~$0$ so that the number of objective chores is a multiple of the number of agents.
Next, Alice and Bob pick those objective chores in a round-robin fashion---Alice picks first, followed by Bob.
Suppose the resulting partial allocation is $(\{o, o_3\}, \{o_2, o_4\})$.
Finally, we reverse the picking order, that is, now, Bob picks first his favourite good from the remaining items and Alice next.
The resulting allocation is $(\{o, o_3\}, \{o_2, o_4, o_1\})$; one can verify that the allocation is \EFOne.

\begin{theorem}[\citet{AzizCaIg22}]
For additive utilities, the double round-robin algorithm returns an \EFOne allocation in polynomial time.
\end{theorem}

In the indivisible-goods setting, another well-known method to compute an \EFOne allocation (for any number of agents with arbitrary monotonic utilities) is the \emph{envy-cycle elimination algorithm} of \citet{LiptonMaMo04}, which works by iteratively allocating a good to an agent who is not envied by anyone else.
We can always find such an agent by resolving \emph{envy cycles} in the underlying envy graph of the partial allocation.

As observed in the work of \citet{BercziBeBo20} and \citet{BhaskarSrVa21}, however, a naive extension of the method to the indivisible-chores setting (even for agents with additive utilities) could fail to find an \EFOne allocation if envy cycles are resolved in an \emph{arbitrary} way, let alone for mixed indivisible goods and chores.
Intuitively speaking, this is because even if an agent gets a better bundle when we resolve an envy cycle, the bundle may not contain a large enough chore whose removal eliminates the envy.
Nevertheless, \citet{BhaskarSrVa21} introduced a key insight that we can always resolve the \emph{top-trading envy cycle}, in which each agent only points to the agent she envies the most, and preserve \EFOne.
Such an insight also works for doubly-monotonic instances.

\begin{theorem}[\citet{BhaskarSrVa21}]
For doubly-monotonic utilities, a modified top-trading envy-cycle elimination algorithm~\citep[see][Algorithm~3]{BhaskarSrVa21} computes an \EFOne allocation.
\end{theorem}

Looking beyond additive utilities, \citet{CousinsViZi23} introduced the class of \emph{order-neutral submodular valuations}, which relaxes the assumption that each item must be classified as a good or a chore (like the assumption in doubly-monotonic utility functions), but comes with a stronger restriction of submodularity.
Further restricting the possible marginal values to~$-1$, $0$, and~$c$ (a positive integer), \citet{CousinsViZi23} showed that a leximin allocation\footnote{A leximin allocation is one that maximizes the minimum among the agents' utilities; subject to this, it maximizes the second smallest utility, and so on.} can be computed efficiently; such an allocation, however, may not be \EFOne even with two agents.
For two agents with arbitrary utility functions over mixed indivisible goods and chores, \citet{BercziBeBo20} devised a polynomial-time algorithm based on the envy graph that always computes an \EFOne allocation.

\begin{problem}
For three (or more) agents with arbitrary utility functions over mixed indivisible goods and chores, does there always exist an \EFOne allocation?
This question remains open even if agents have identical utility functions.
\end{problem}

What about additionally demanding Pareto optimality?
The double round-robin and the modified top-trading envy-cycle elimination methods return allocations that are \EFOne but may not be PO.
In the context of allocating goods alone and additive utilities, the \emph{maximum Nash welfare (MNW)} allocation satisfies both \EFOne and PO~\citep{CaragiannisKuMo19}.\footnote{With indivisible goods, an MNW allocation deals with the ``drowning by zero'' problem by first maximizing the number of agents receiving positive utilities, and then maximizing the product of these positive utilities.}
The question regarding whether an \EFOne and PO allocation always exists for indivisible chores alone remains unresolved, except for the cases of up to three additive agents~\citep{AzizCaIg22,GargMuQi23},\footnote{\citet{GargMuQi23}'s result holds for \EFOne and fPO.
An allocation is said to satisfy \emph{fractional Pareto optimality (fPO)} if it is not Pareto dominated by any fractional allocation, in which an agent may receive a fractional share of an \emph{indivisible} good~\citep{BarmanKrVa18}.} bi-valued instances~\citep{EbadianPeSh22,GargMuQi22}, and two types of chores~\citep{AzizLiRi23}.
For two agents with additive utilities over mixed indivisible goods and chores, \citet{AzizCaIg22} showed that an \EFOne and PO allocation can always be found using a discrete version of the well-known Adjusted Winner (AW) rule~\citep{BramsTa96}.
A natural question is whether we can extend this to three (or more) agents.

\begin{problem}
With mixed indivisible goods and chores, for three (or more) agents and additive utilities, does an \EFOne and PO allocation always exist?
Recall that this question remains open even in the indivisible-chores setting.

If so, can we compute the allocation in polynomial time?
Note that it remains unknown whether, in the indivisible-goods setting, an \EFOne and PO allocation can be computed in polynomial time.
\end{problem}

When weakening \EFOne to PROP1, the existence and computation of a PROP1 and PO allocation has been resolved by \citet{AzizMoSa20}, even if agents have unequal entitlements.\footnote{We refer the interested readers to the recent review by \citet{Suksompong25}, which discussed about fair division involving agents with unequal entitlements.}

\begin{theorem}[\citet{AzizMoSa20}]
For additive utilities over indivisible goods and chores, there exists a polynomial-time algorithm that always computes a PROP1 and PO allocation.
\end{theorem}

So far, we have been only concerned with notions of individual fairness.
Inspired by the concept of \emph{group envy-freeness (GEF)}~\citep{BerliantThDu92}---a generalization of envy-freeness for equal-sized groups of agents,\footnote{An allocation~$(\bundle_i)_{i \in N}$ is said to satisfy \emph{group envy-freeness (GEF)} if for every non-empty groups of agents~$S, T \subseteq N$ with $|S| = |T|$, there is no reallocation~$(B_i)_{i \in S}$ of resources~$\bigcup_{i \in T} \bundle_i$ among agents~$S$ such that for every~$i \in S$, $u_i(B_i) \geq u_i(\bundle_i)$, with one strict inequality.} \citet{AzizRe20} formalized relaxations of GEF for the case of mixed indivisible goods and chores.
We include their ``up to one'' relaxation here.
An allocation~$(\indivisibleItems_i)_{i \in N}$ of indivisible items~$\indivisibleItems$ is said to satisfy \emph{GEF up to one item (GEF1)} if for every non-empty groups of agents~$S, T \subseteq N$ with $|S| = |T|$ and every reallocation~$(B_i)_{i \in S}$ of items~$\bigcup_{i \in T} \indivisibleItems_i$ among agents~$S$, there exists an item~$o_i \in (\indivisibleItems_i \cap \indivisibleChores_i) \cup (B_i \cap \indivisibleGoods_i)$ for each~$i \in S$ such that $(B_i \setminus \{o_i\})_{i \in S}$ does not Pareto dominate $(\indivisibleItems_i \setminus \{o_i\})_{i \in S}$.
\citet{AzizRe20} devised polynomial-time algorithms to compute a GEF1 allocation when agents have identical utilities, or when agents have ternary symmetric utilities of the form~$\{-\alpha_i, 0, \alpha_i\}$ for a given~$\alpha_i > 0$.

What if we consider a stronger fairness property like \EFX?
With additive utilities, \EFX allocations do not always exist.
This can be seen from an instance with a mixture of \emph{objective} goods and chores and \emph{lexicographic preferences}~\citep{HosseiniSiVa23}.\footnote{Let~$\mathcal{L}$ be set of all (strict and complete) linear orders over items~$\indivisibleItems$.
Denote by~$\rhd \coloneqq (\rhd_1, \rhd_2, \dots, \rhd_n)$ the \emph{importance profile} that specifies for each agent~$i \in N$ an \emph{importance ordering}~$\rhd_i \in \mathcal{L}$ over~$\indivisibleItems$.
Given any two non-identical bundles~$X$ and~$Y$, let~$z \in (X \setminus Y) \cup (Y \setminus X)$ be the most important item according to~$\rhd_i$.
Lexicographic preferences say that agent~$i$ prefers bundle~$X$ over bundle~$Y$ if either~$z \in X \cap \indivisibleGoods_i$ or~$z \in Y \cap \indivisibleChores_i$.
Lexicographic preferences can be seen a special case of additive utilities in which the magnitude of utilities grow exponentially in the importance ordering.}
However, for special classes of indivisible goods and chores such as absolute identical utilities (i.e., for each item, the agents' utilities have identical magnitudes but may have different signs), ternary utilities of the form $\{\alpha, 0, -\beta\}$, or separable lexicographic preferences (i.e., either chores are more important than goods or goods than chores), there exist polynomial-time algorithms that always return an \EFX and PO allocation~\citep{AleksandrovWa20-KI,HosseiniMaWa23}.
With non-additive utilities, we refer interested readers to the work of \citet{BercziBeBo20} for various ways of defining \EFX and their (non-)existence results.

\begin{problem}
Are there other natural subclasses of additive utilities over mixed indivisible goods and chores that always admit an \EFX allocation?
Or even an \EFX and PO allocation?
\end{problem}

We remark that the question is of interest even if we only consider indivisible goods or chores.
It remains unknown whether there always exists an \EFX allocation of indivisible goods (resp., chores) for at least four (resp., three) agents with additive valuations~\citep{ChaudhuryGaMe24,ChristoforidisSa24,ZhouWu24}.

\subsection{MMS}
\label{sec:indivisible-items:MMS}

Given \Cref{def:MMS}, the most natural and intriguing question is whether an MMS allocation always exists.
The seminal work of \citet{KurokawaPrWa18} showed that, with only indivisible goods, an MMS allocation may not exist when there are at least three agents, but $\frac{2}{3}$-MMS can always be satisfied.
Since then, many subsequent works have been carried out on improving the approximation ratio, designing simpler algorithms or giving simpler analyses, considering more general valuations, studying the indivisible-chores setting, etc.
We refer interested readers to Section~5 of \citet{AmanatidisAzBi23} and Section~7.1 of \citet{GuoLiDe23} for a detailed account of recent developments on computing approximate-MMS allocations in the indivisible-goods and indivisible-chores settings, respectively.

In what follows, we mainly focus on the developments in the setting where we allocate mixed indivisible goods and chores.
We start by discussing about the computation of agents' MMS values.
It is well-known that an agent's maximin share is NP-hard to compute, even with only indivisible goods~\citep[see, e.g.,][]{KurokawaPrWa18}.
Nevertheless, with indivisible goods, there exists a \emph{polynomial-time approximation scheme (PTAS)} to approximate each agent's maximin share~\citep{Woeginger97}.
To be more precise, given a constant~$\varepsilon > 0$, we can compute in polynomial time a partition $(P_1, P_2, \dots, P_n)$ of the set of indivisible goods~$\resource$ for agent~$i$ such that
\[
\min_{j \in [n]} u_i(P_j) \geq (1 - \varepsilon) \cdot \MMS_i(n, \resource).
\]
Furthermore, there exist polynomial-time approximation schemes to approximate an agent's maximin share when allocating indivisible chores~\citep[e.g.,][]{JansenKlVe20}, or mixed divisible and indivisible goods~\citep{BeiLiLu21}.

With mixed indivisible goods and chores, however, computing an approximate MMS value is more challenging.
\citet{KulkarniMeTa21} showed that it is NP-hard to approximate an agent's MMS value up to any approximation factor in~$(0, 1]$.
Intuitively speaking, the bottleneck is that the absolute value of MMS can be arbitrarily small (or, in other words, an MMS value can be arbitrarily close to~$0$).
\citet{KulkarniMeTa21-AAAI} later gave a PTAS to compute an agent's MMS value when its absolute value is at least~$1/p$ times either the total value of all the goods or total cost of all the chores, for some constant~$p$ greater than~$1$.

We now discuss to what extent we can compute an approximate-MMS allocation.
Note that in both indivisible-goods and indivisible-chores settings, a constant approximation exists.\footnote{The state-of-the-art approximation ratio is $\frac{3}{4} + \frac{3}{3836}$ for goods due to \citet{AkramiGa24} and $\frac{11}{13}$ for chores due to \citet{HuangSe23}.
We remark that the factor of~$\frac{11}{13}$, instead of~$\frac{13}{11}$ in~\citep{HuangSe23}, is due to the fact that we assume agents have non-positive values for chores while \citet{HuangSe23} (and almost all of the works on approximate-MMS allocations of indivisible chores) assume (non-negative) \emph{cost} functions for the agents.}
In contrast, with mixed indivisible goods and chores, for any fixed~$\alpha \in (0, 1]$, an $\alpha$-MMS allocation may not exist~\citep{KulkarniMeTa21}.
And since the problem of finding an $\alpha$-MMS allocation is NP-hard for any~$\alpha \in (0, 1]$, \citet{KulkarniMeTa21} approached the problem by designing computationally efficient algorithms, which, given a mixed-items fair division instance and $\alpha, \varepsilon \in (0, 1]$, can compute an $(\alpha - \varepsilon)$-MMS allocation (in addition to being approximately PO) of the given instance, or report that no $\alpha$-MMS allocation exists for the instance.
Note that their algorithms hinge upon certain conditions regarding the instances and thus only work for a subclass of instances satisfying the specified conditions.
Specifically, for the special case of a constant number of agents where the total value of goods is some factor away of the total absolute value of chores, \citet{KulkarniMeTa21} gave a PTAS to find an $(\alpha - \varepsilon)$-MMS and $\gamma$-PO allocation when given $\varepsilon, \gamma > 0$, for the highest possible $\alpha \in (0, 1]$.
Along the way, they developed a novel approach of using an LP-rounding through envy-cycle elimination as a tool to ensure PO with $\alpha$-MMS.

The aforementioned works motivate the study of computing (approximate-)MMS (and possibly with PO) allocations if agents' preferences are more restricted.
To this end, given lexicographic preferences over mixed indivisible goods and chores, an MMS and PO allocation always exists and can be computed in polynomial time~\citep{HosseiniSiVa23,HosseiniMaWa23}.

\subsection{Further Work}

Starting with the work of \citet{BogomolnaiaMoSa17}, a line of research has addressed the fair allocation of mixed homogeneous \emph{divisible} goods and chores~\citep{GargMc20,ChaudhuryGaMc23,GargHoMc21}, focusing on a central solution concept in economics called \emph{competitive equilibrium}~\citep{ArrowDe54}.
\citet{Segal-Halevi18} considered the fair division of a \emph{heterogeneous} divisible resource that contains both good parts and bad parts, and proved that a connected envy-free division of the resource always exists for three agents.
Later, \citet{MeunierZe19} extended the existence of a connected envy-free division to the case where~$n$ is a prime number or $n = 4$.

Such divisible allocations of goods and chores might be adapted into randomized algorithms for indivisible goods and chores.
This then naturally suggests another interesting direction for future study: algorithm design for mixed indivisible goods and chores with good ex-ante \emph{and} ex-post properties.
Such a ``best-of-both-worlds'' perspective has recently been receiving attention when allocating indivisible goods~\citep{AkramiGaSh24,AkramiMeSe23,AzizFrSh23,AzizGaMi23,BabaioffEzFe22,FeldmanMaNa24,HoeferScVa23} and in \emph{collective choice} contexts~\citep{AzizLuSu23,AzizLuSu24,SuzukiVo24}.

\begin{problem}
Can we obtain a randomized allocation of mixed indivisible goods and chores which has good (exact) fairness ex ante from which we can construct integral allocations with good (approximate) fairness ex post?
\end{problem}

We conclude this section by pointing out studies which generalize the mixed indivisible goods and chores setting.
For instance, \citet{CaragiannisNa24} studied a repeated matching setting where a set of items is matched to the same set of agents repeatedly over multiple rounds.
In their model, each agent gets exactly one item per round, and her value for the item depends on how many times she has matched to the item in the previous rounds and can be positive, zero or negative.
Among other results, \citet{CaragiannisNa24} showed that with mixed items, a matching that is \emph{envy-free up to one swap} exists for identical agents and in several other special cases if agents have heterogeneous valuations.
In this survey, we assume that agents have preferences over the items, but not the other way around.
\citet{IgarashiKaSu23} studied a fair division setting with \emph{two-sided preferences}~\citep[see also][]{FreemanMiSh21}, that is, additionally, the items also have preferences over the agents.
They focused on guaranteeing \EFOne for the agents together with a stability condition for both sides.
Some of their results allow the utilities to be either positive or negative.
Again, we assume in this survey that agents only derive utilities from their own received items.
Other work (such as \citealp{BranzeiPrZh13,LiZhZh15,SeddighinSaGh21,AzizSuSu23}) have considered fair division with \emph{externalities} in which each agent also receives (positive or negative) utilities from items that are assigned to other agents.

\section{Mixed Divisible and Indivisible Goods}
\label{sec:mixed-div-ind}

This section is concerned with the fair division of mixed divisible and indivisible goods described in \Cref{sec:prelim:mixed-div-ind}.
We will first focus on how to obtain approximately envy-free allocations in \Cref{sec:mixed-div-ind:EFM} and next turn our attention to allocations guaranteeing agents their fair share (depending on how we define it) in \Cref{sec:mixed-div-ind:MMS}.

\subsection{Envy-freeness Relaxations}
\label{sec:mixed-div-ind:EFM}

When allocating mixed goods, \citet{BeiLiLi21} proposed the following fairness concept called \emph{envy-freeness for mixed goods} that naturally generalizes envy-freeness and \EFOne to the mixed-goods model and is guaranteed to exist.

\begin{definition}[\EFMzero~{\citep[Definition~2.3]{BeiLiLi21}}]
\label{def:EFMzero}
An allocation $\alloc = (\bundle_i)_{i \in N}$ of mixed goods~$\resource = \divisibleItems \cup \indivisibleItems$ is said to satisfy \emph{envy-freeness for mixed goods (\EFMzero)} if for any pair of agents~$i, j \in N$,
\begin{itemize}
\item if agent~$j$'s bundle~$\bundle_j$ consists of only indivisible goods, there exists some good~$g \in \bundle_j$ such that $u_i(\bundle_i) \geq u_i(\bundle_j \setminus \{g\})$;
\item otherwise, $u_i(\bundle_i) \geq u_i(\bundle_j)$.
\end{itemize}
\end{definition}

At a high level, \EFMzero requires that an agent is envy-free towards any agent whose bundle contains \emph{a positive amount} of divisible resources and \EFOne towards the rest.
It can be verified that with only divisible (resp., indivisible) goods, \EFMzero reduces to envy-freeness (resp., \EFOne).
Moreover, an \EFMzero allocation of mixed goods always exists.

\begin{theorem}[\citet{BeiLiLi21}]
\label{thm:EFM-zero-exists-goods}
An \EFMzero allocation of mixed goods always exists for any number of agents and can be found in polynomial time with polynomially many \RW queries and calls to an oracle which could return a perfect partition of a cake.
\end{theorem}

The high-level algorithmic idea to compute an \EFMzero allocation is as follows:
\begin{itemize}
\item We start with an \EFOne allocation of the indivisible items.
The partial allocation is therefore \EFMzero.
(The \EFMzero property will be an invariant of the algorithm.)

\item Next, we construct an envy graph~$(N, E_{\text{envy}} \cup E_{\text{eq}})$ for the partial allocation, where each vertex in the envy graph corresponds to an agent, and~$E_{\text{envy}}$ and~$E_{\text{eq}}$ consist of the following two types of edges, respectively:
\begin{itemize}
\item if $u_i(\bundle_i) < u_i(\bundle_j)$, we establish an \emph{envy edge} from~$i$ to~$j$, i.e., $(i, j) \in E_{\text{envy}}$;
\item if $u_i(\bundle_i) = u_i(\bundle_j)$, we establish an \emph{equality edge} from~$i$ to~$j$, i.e., $(i, j) \in E_{\text{eq}}$.
\end{itemize}
A cycle in an envy graph is called an \emph{envy cycle} if it contains at least one envy edge.
Given an envy graph, a non-empty subset of agents~$S \in N$ forms an \emph{addable set} if
\begin{itemize}
\item there is no envy edge between any pair of agents in~$S$;
\item there is no edge from any agent in~$N \setminus S$ to any agent in~$S$.
\end{itemize}

\item Then, we identify a maximal addable set among whom we divide some divisible resources using a perfect allocation~\citep{Alon87} --- we ensure that the \EFMzero property is still preserved.
Along the way, in order to identify an addable set, we may need to rotate bundles of the agents involved in an envy cycle.
This step is repeated until we allocate all divisible resources.
\end{itemize}

A challenge is that the perfect allocation cannot be implemented with a finite number of queries in the RW query model, even if there are only two agents~\citep{RobertsonWe98}.
Nevertheless, an \EFMzero (and hence \EFM) allocation can be computed efficiently for two agents with general additive valuations and for $n$ agents with \emph{piecewise linear} density functions over the cake~\citep{BeiLiLi21}.

\begin{problem}
\label{problem:bounded-EFM-alg}
Does there exist a bounded or even \emph{finite} protocol in the RW query model to compute an \EFM allocation?
\end{problem}

Despite the strong fairness guarantee provided by \EFMzero, the notion is incompatible with PO \citep[Example~6.3]{BeiLiLi21}.
The counter-example hinges on the fact that in an \EFMzero allocation, agent~$i$ should not envy agent~$j$ if agent~$j$'s bundle contains any positive \emph{amount} of the cake, although agent~$i$ may value the piece of cake at~$0$.
In the original paper of \citet{BeiLiLi21}, the fairness criterion is simply called \EFM; we rename it by following the nomenclature of \citet{KyropoulouSuVo20} for \EFXzero and \EFX (cf.\ \Cref{ft:EFX0-EFX-nomenclature}).
We let \EFM be the shorthand for a more natural variant defined below.

\begin{definition}[\EFM~{\citep[Definition~6.4]{BeiLiLi21}}]
\label{def:EFM-goods}
An allocation $\alloc = (\bundle_i)_{i \in N}$ of mixed goods~$\resource = \divisibleItems \cup \indivisibleItems$ is said to satisfy \emph{weak envy-freeness for mixed goods (\EFM)} if for any pair of agents~$i, j \in N$,
\begin{itemize}
\item if agent~$j$'s bundle consists of indivisible goods with either no divisible good or divisible good that yields value~$0$ to agent~$i$ (i.e., $u_i(\divisibleItems_j) = 0$), there exists an indivisible good~$g \in \bundle_j$ such that $u_i(\bundle_i) \geq u_i(\bundle_j \setminus \{g\})$;
\item otherwise, $u_i(\bundle_i) \geq u_i(\bundle_j)$.
\end{itemize}
\end{definition}

A strengthening of \EFMzero is to incorporate the idea of being \EFXzero when comparing to a bundle with only indivisible goods~\citep[see, e.g.,][]{BeiLiLi21,NishimuraSu23}.\footnote{The notion can also be refined by using the \EFX criterion (\Cref{def:EFX_0-EFX-goods}).
For the purpose of this survey, we do not explicitly give its definition here.}
An allocation~$\alloc = (\bundle_1, \bundle_2, \dots, \bundle_n)$ of mixed goods~$\resource = \divisibleItems \cup \indivisibleItems$ is said to satisfy \emph{envy-freeness up to any good for mixed goods (\EFXM)} if for any pair of agents~$i, j \in N$,
\begin{itemize}
\item if agent~$j$'s bundle consists of only indivisible goods, $u_i(\bundle_i) \geq u_i(\bundle_j \setminus \{g\})$ for \emph{any} (indivisible) good~$g \in \bundle_j$;
\item otherwise, $u_i(\bundle_i) \geq u_i(\bundle_j)$.
\end{itemize}

It follows from the definitions that \EF{} $\implies$ \EFXM $\implies$ \EFMzero $\implies$ \EFM.
Given any mixed-goods instance, if an \EFXzero allocation of indivisible goods exists, we can start with this \EFXzero allocation, apply the rest of the above \EFMzero algorithmic framework, and eventually compute an \EFXM allocation of the mixed-goods instance.

We demonstrate \EFXM, \EFMzero and \EFM allocations below.

\begin{example}
Consider a mixed-good instance with three indivisible goods~$\{g_1, g_2, g_3\}$, one homogeneous divisible good~$D$, two agents and their valuations as follows:

\begin{center}
\begin{tabular}{@{}c|*{3}{c}|c@{}}
\toprule
& $g_1$ & $g_2$ & $g_3$ & $D$ \\
\midrule
$u_1$ & $2$ & $1$ & $1$ & $0$ \\
$u_2$ & $2$ & $1$ & $1$ & $1$ \\
\bottomrule
\end{tabular}
\end{center}

Let us consider the following three allocations:
\begin{center}
\begin{tabular}{@{}l|*{2}{l}@{}}
\toprule
& Agent~$1$ & Agent~$2$ \\
\midrule
Allocation~$\alloc$ & $\{g_1, g_2\}$ & $\{g_3, D\}$ \\
Allocation~$\alloc'$ & $\{g_3, D\}$ & $\{g_1, g_2\}$ \\
Allocation~$\alloc''$ & $\{g_3\}$ & $\{g_1, g_2, D\}$ \\
\bottomrule
\end{tabular}
\end{center}

Allocation~$\alloc$ is \EFXM (but not envy-free), because $a_2$ envies $a_1$, but the envy can be eliminated by removing~$a_1$'s least valued good (i.e., good~$g_2$) from $a_2$'s bundle.

Allocation~$\alloc'$ is \EFMzero (but not \EFXM), because
\begin{itemize}
\item $u_1(\{g_3, D\}) = 1 \geq u_1(\{g_1, g_2\} \setminus \{g_1\})$ (showing \EFMzero);
\item $u_1(\{g_3, D\}) = 1 < 2 = u_1(\{g_1, g_2\} \setminus \{g_2\})$ (failing \EFXM).
\end{itemize}

Allocation~$\alloc''$ is not \EFMzero, because $a_2$'s bundle contains divisible good~$D$, yet still $a_1$ envies~$a_2$.
The allocation, however, is \EFM.
As $a_1$ values the divisible good at~$0$, according to \Cref{def:EFM-goods}, we only need to examine whether $a_1$'s envy towards $a_2$ can be eliminated by removing an indivisible good from $a_2$'s bundle.
And indeed this is the case since $u_1(\{g_3\}) = 1 = u_1(\{g_1, g_2, D\} \setminus \{g_1\})$.
\end{example}

We introduce here the two variants, \EFMzero and \EFM, as both notions have their own merits.
On the one hand, \EFMzero is conceptually easier to be strengthened or extended when considering more general settings, e.g., with non-additive utilities,\footnote{With indivisible items, \citet{BercziBeBo20} already discussed several ways to extend \EFX when agents have non-additive utilities.} and any existence result of \EFMzero may still be carried over to \EFM (if well-defined).
On the other hand, \EFM precludes the counter-intuitive incompatibility with PO~\citep[Example~6.3]{BeiLiLi21}.
However, \EFM is incompatible with fPO~\citep{BeiLiLi21}.
The compatibility between \EFM and PO is still unresolved and is an very interesting open question.

\begin{problem}
\label{problem:EFM-PO}
Are \EFM and PO compatible?
\end{problem}

Despite providing strong compatibility between PO and (approximate) envy-freeness, the maximum Nash welfare (MNW) allocation fails to guarantee a PO and EFM allocation given mixed goods~\citep{BeiLiLi21}.
Nevertheless, \citet{NishimuraSu23} provided a formal proof showing that an MNW allocation for mixed goods is PO and \emph{envy-free up to one indivisible good for mixed goods (\EFoneM)}~\citep{CaragiannisKuMo19}, which is based on the idea of removing an indivisible good from an envied bundle to eliminate envy and is weaker than \EFM.
When restricting agents' utilities to binary and linear, an MNW allocation is PO and \EFXM~\citep{NishimuraSu23}.

\citet{BertsimasFaTr11} and \citet{CaragiannisKaKa12} introduced independently the concept of \emph{price of fairness} for quantifying the efficiency loss due to fairness requirements.
Taking \EFMzero as an example, the \emph{price of \EFMzero} is the worst-case ratio between the total utility under an (unconstrained) optimal allocation, and the total utility under an optimal \EFMzero allocation.
Since then, a series of follow-up research has provided tight (for two agents) or asymptotically tight (for~$n$ agents) bounds on the price of approximate-EF notions (like \EFOne, \EFXzero, \EFMzero and \EFXM) when agents have scaled (alternatively, normalized) or unscaled utilities~\citep{BarmanBhSh20,BeiLuMa21,BuLiLi23,LiLiLu24}.
Other questions concerning simultaneously fairness and economic efficiency, for example, maximizing social welfare within fair allocations~\citep{AzizHuMa23,BeiChHu12,BuLiLi23,CohlerLaPa11,SunChDo23}, are equally relevant and worthy of exploration in mixed fair division settings.

While \Cref{thm:EFM-zero-exists-goods} was presented in the context of additive utilities, neither the algorithm of \citet{BeiLiLi21} nor its analysis hinges on the assumption of the utilities over indivisible goods being additive.
As a matter of fact, \EFMzero (and hence \EFM) can still always be satisfied even if agents have monotonic utilities over the indivisible goods, as long as (i) agents' utilities over the divisible goods are additive and (ii) agents' utilities across divisible and indivisible goods are additive.
Below, we give two examples showing that if either condition~(i) or~(ii) is violated, an \EFM allocation may not exist.
Given an interval~$[a, b]$, denote its length as~$\length{[a, b]} = b - a$.
Let~$\widehat{\divisibleItems}$ be a piece of cake consisting of a set of intervals~$\mathcal{I}_{\widehat{\divisibleItems}}$.
Then, $\length{\widehat{\divisibleItems}} = \sum_{I \in \mathcal{I}_{\widehat{\divisibleItems}}} \length{I}$ is the length of the piece of cake~$\widehat{\divisibleItems}$.
Let~$\varepsilon$ be an arbitrarily small positive number.

\begin{example}
This example will show that an \EFM allocation may not exist if agents' utilities over divisible goods are not additive.
Consider two agents dividing an indivisible good~$g$ and a divisible good~$D = [0, 1]$.
Both agents have identical utility function~$u$, where $u(g) = 1$ and
\[
u(\widehat{\divisibleItems}) =
\begin{cases}
1 + \varepsilon & \text{if}~\frac{1}{2} + \varepsilon \leq \length{\widehat{\divisibleItems}}; \\
\frac{\varepsilon}{2} & \text{if}~0 < \length{\widehat{\divisibleItems}} < \frac{1}{2} + \varepsilon; \\
0 & \text{if}~\length{\widehat{\divisibleItems}} = 0.
\end{cases}
\]
We have $u(\{g\} \cup \widehat{\divisibleItems}) = u(g) + u(\widehat{\divisibleItems})$, i.e., agents' utilities across divisible and indivisible goods are additive.
Assume without loss of generality that agent~$1$ gets good~$g$.
We distinguish the following two cases and show that in either case, the allocation is not \EFM.
\begin{itemize}
\item $\length{D_2} \geq \frac{1}{2} + \varepsilon$: Agent~$2$ has divisible good that is positively valued by agent~$1$; however, because $u(\{g\} \cup D_1) \leq 1 + \frac{\varepsilon}{2} < 1 + \varepsilon = u(D_2)$, agent~$1$ envies agent~$2$.
\item $\length{D_2} < \frac{1}{2} + \varepsilon$: Agent~$1$ has divisible good that is positively valued by agent~$2$; however, because $u(D_2) \leq \frac{\varepsilon}{2} < 1 \leq u(\{g\} \cup D_1)$, agent~$2$ envies agent~$1$.
\end{itemize}
\end{example}

\begin{example}
This example will show that an \EFM allocation may not exist if agents' utilities across divisible and indivisible goods are not additive.
Consider two agents dividing an indivisible good~$g$ and a homogeneous divisible good~$\divisibleItems = [0, 1]$.
They have identical utility function~$u$ where $u(g) = \frac{1}{2} - \varepsilon$, $u(\widehat{\divisibleItems}) = \length{\widehat{\divisibleItems}}$, and
\begin{equation*}
u(\{g\} \cup \widehat{\divisibleItems}) =
\begin{cases}
u(g) + u(\widehat{\divisibleItems}) & \text{if}~\length{\widehat{\divisibleItems}} \geq \frac{1}{2}; \\
\max\{u(g), u(\widehat{\divisibleItems})\} & \text{if}~0 \leq \length{\widehat{\divisibleItems}} < \frac{1}{2}.
\end{cases}
\end{equation*}
Assume without loss of generality that agent~$1$ gets good~$g$.
We distinguish the following two cases and show that in neither case, the allocation is \EFM.
\begin{itemize}
\item $\length{\divisibleItems_2} > \frac{1}{2}$: Agent~$2$ has divisible good that is positively valued by agent~$1$; however, because $u(\{g\} \cup D_1) = \max\{u(g), u(D_1)\} < \frac{1}{2} < u(D_2)$, agent~$1$ envies agent~$2$.
\item $\length{\divisibleItems_2} \leq \frac{1}{2}$: Agent~$1$ has divisible good that is positively valued by agent~$2$; however, because $u(D_2) \leq \frac{1}{2} < 1 - \varepsilon \leq u(\{g\} \cup D_1)$, agent~$2$ envies agent~$1$.
\end{itemize}
\end{example}

\citet{BhaskarSrVa21} studied an extension of the mixed-goods model as follows.
In their \emph{mixed-resources} model, the resource~$\resource$ consists of a set~$\indivisibleItems = [m]$ of indivisible \emph{items} as defined in \Cref{sec:prelim:indivisible-items} and a divisible resource~$[0, 1]$ which is either an objective divisible good (i.e., $\forall i \in N$, $f_i \colon [0, 1] \to \mathbb{R}_{\geq 0}$) or an objective divisible chore (i.e., $\forall i \in N$, $f_i \colon [0, 1] \to \mathbb{R}_{\leq 0}$), referred to as a ``bad cake'' by \citet{BhaskarSrVa21}.
An allocation of the mixed resources and agents' utilities in the allocation are defined the same way as in \Cref{sec:prelim:mixed-div-ind}.
\citeauthor{BhaskarSrVa21} extended the formulation of \EFM as follows.

\begin{definition}[\EFM for mixed resources~\citep{BhaskarSrVa21}]
In the mixed-resources model, an allocation $\alloc = (\bundle_1, \bundle_2, \dots, \bundle_n)$ is said to satisfy \emph{envy-freeness for mixed resources (\EFM)} if for any pair of agents~$i, j \in N$, either~$i$ does not envy~$j$, that is, $u_i(\bundle_i) \geq u_i(\bundle_j)$, or all of the following hold:
\begin{itemize}
\item $u_i(\divisibleItems_i) \geq 0$, i.e., $i$ does not have any bad cake,
\item $u_i(\divisibleItems_j) \leq 0$, i.e., $j$ does not have any cake, and
\item $\exists o \in \indivisibleItems_i \cup \indivisibleItems_j$ such that $u_i(\bundle_i \setminus \{o\}) \geq u_i(\bundle_j \setminus \{o\})$.
\end{itemize}
\end{definition}

\begin{theorem}[\citet{BhaskarSrVa21}]
An \EFM allocation always exists when allocating mixed resources consisting of doubly-monotonic indivisible items and a divisible chore.
\end{theorem}

The algorithmic framework introduced earlier to obtain an \EFMzero allocation does not seem to work when allocating indivisible chores and a cake~\citep{BhaskarSrVa21}.
In special cases where agents have identical rankings of the indivisible chores or $m \leq n+1$, \citet{BhaskarSrVa21} proved the existence of an \EFM allocation.

\begin{problem}
Does there always exist an \EFM allocation when allocating indivisible chores and a cake?
\end{problem}

An affirmative answer to the above question may pave the way for solving the existence of \EFM in a more general setting where resource~$\resource$ consists of divisible and indivisible items, and each item, either divisible or indivisible, may be a good to some agents but a chore for others.

As valuations are elicited from the agents, the power and limitations of \emph{truthful} mechanisms in addition to being fair have been explored in a variety of resource allocation scenarios~\citep[see, e.g.,][]{BeiLuSu24,BogomolnaiaMo04,BrandlBrPe21,FreemanPePe21,FreemanSc24,FriedmanGkPs19,LiZhZh15,ViswanathanZi23}.
Truthfulness (or \emph{strategyproofness}) requires that it should be in every agent's best interest to report her true underlying preferences to the mechanism.

For instance, in cake cutting, \citet{ChenLaPa13} designed a truthful envy-free mechanism for agents with piecewise-uniform valuations when assuming \emph{free disposal}, which means that the mechanism is allowed to throw away part of the resources at no cost.
\citet{BeiHuSu20} then removed the free disposal assumption and exhibited truthful envy-free cake cutting mechanisms for two agents with piecewise-uniform valuations as well as for multiple agents with more restricted classes of valuations.
\citet{BuSoTa23} later showed that for piecewise-constant valuations, there does not exist a truthful proportional cake cutting mechanism.

Moving to indivisible-goods setting, truthfulness and \EFOne are incompatible for two agents with additive valuations~\citep{AmanatidisBiCh17}.
Nevertheless, for \emph{binary additive} valuations, \citet{HalpernPrPs20} showed the MNW rule with lexicographic tie-breaking is \EFOne, PO and \emph{group strategyproof} (no coalition of agents can misreport their preferences in a way that they all benefit).
Concurrently and independently, for \emph{binary submodular} (also known as \emph{matroid-rank}) valuations, i.e., valuations are submodular functions with binary marginals, \citet{BabaioffEzFe21} designed a mechanism that is truthful and returns an \EFOne and PO allocation.
Their mechanism was then proved to be group strategyproof by \citet{BarmanVe22}.
Those results have also been generalized to the setting where agents have unequal entitlements~\citep{SuksompongTe22,SuksompongTe23}.

Regarding a mixture of both divisible and indivisible goods, \citet{LiLiLu23} modelled the mixed goods as a set of indivisible goods together with a set of \emph{homogeneous} divisible goods.
While truthfulness and \EFM are incompatible even if there are only two agents having additive utilities over a single indivisible good and a single divisible good, they designed truthful and \EFM mechanisms in several special cases where the expressiveness of agents' utilities are further restricted.

\begin{problem}
An intriguing question left open in \citep{LiLiLu23} is to show the (in)compatibility between truthfulness and \EFM when $n \geq 3$ agents have binary additive utilities over an arbitrary number of indivisible and divisible goods.
\end{problem}

We remark that as an \EFoneM allocation of mixed goods can be obtained by combining an \EFOne allocation of the indivisible goods and an envy-free allocation of the divisible goods, a truthful \EFoneM mechanism can be obtained by combining a truthful \EFOne mechanism (for indivisible goods) and a truthful envy-free mechanism (for divisible goods).

\subsection{MMS and \texorpdfstring{PROP-$\alpha$}{PROP-alpha}}
\label{sec:mixed-div-ind:MMS}

We have seen that the MMS guarantee has been extensively studied for indivisible items, and the notion is well-defined in the mixed-goods model, to which \citet{BeiLiLu21} extended the study of (approximate) MMS guarantee.

Given a mixed-goods instance, let the \emph{MMS approximation guarantee of the instance} denote the maximum value of~$\alpha$ such that the instance admits an $\alpha$-MMS allocation.
\citet{BeiLiLu21} showed that the worst-case MMS approximation guarantee across all mixed-goods instances is the same as that across all indivisible-goods instances.
It is not surprising as the non-existence of an MMS allocation only arises when the resources to be allocated become indivisible.
This intuition, however, no longer holds for some specific instances.
There exists some instance to which a small amount of divisible goods is added; the MMS approximation guarantee of the new instance strictly decreases.

Concerning the existence and computation of approximate MMS allocations, \citet{BeiLiLu21} devised an algorithm that always produces an $\alpha$-MMS allocation, where~$\alpha$ monotonically increases in terms of the ratio between agents' values for the entire divisible goods and their own maximin share.

\begin{theorem}[\citet{BeiLiLu21}]
Given any mixed-goods instance, an $\alpha$-MMS allocation always exists, where
\[
\alpha = \min \left\{ 1, \frac{1}{2} + \min_{i \in N} \left\{ \frac{u_i(D)}{2 (n-1) \cdot \MMS_i} \right\} \right\}.
\]
\end{theorem}

And even though \citet{BeiLiLu21} discussed an approach to improve the approximation guarantee of their algorithm and can match the state-of-the-art approximation ratio of~$\frac{3}{4} + \frac{3}{3836}$ for indivisible goods due to \citet{AkramiGa24}, improving the ratio further is an interesting future work.

They also discussed how to convert the algorithm into a polynomial-time algorithm at the cost of a small loss in the MMS approximation ratio.
This is achieved by plugging in agents' approximate MMS values.
To be more specific, by using the PTAS of \citet{Woeginger97}, \citet{BeiLiLu21} designed a new PTAS that, given a constant~$\varepsilon > 0$, can compute a partition~$(P_i)_{i \in [n]}$ of mixed goods~$\resource$ for agent~$i$ in polynomial time, such that
\[
\min_{j \in [n]} u_i(P_j) \geq (1 - \varepsilon) \cdot \MMS_i(n, \resource).
\]

Recently, \citet{LiLiLi24} introduced another share-based fairness notion called \emph{proportionality up to $\alpha$-fraction of one good (PROP-$\alpha$)}, which generalizes proportionality and PROP1 to the mixed-good setting.
The core idea behind PROP-$\alpha$ is to refine PROP1 by quantifying the contribution of divisible goods to achieving fairness.
Following this high-level idea, PROP-$\alpha$ directly strengthens the ``up to one'' relaxation to the ``up to a fraction'', where the specific fraction depends on the proportion of indivisible goods relative to all goods.
Intuitively, an agent may desire fairer allocations when share of divisible goods is more valuable. The formal definition of PROP-$\alpha$ can be found as follows.

\begin{definition}[PROP-$\alpha$~\citep{LiLiLi24}]
\label{def:PROPalpha}
An allocation $(\bundle_i)_{i \in N}$ of mixed goods~$\resource = \divisibleItems \cup \indivisibleItems$ is said to satisfy \emph{proportionality up to $\alpha$-fraction of one good (PROP-$\alpha$)} if for any agent~$i \in N$, there exists an indivisible good~$g \in \indivisibleItems \setminus \bundle_i$ such that $u_i(\bundle_i) + \alpha_i \cdot u_i(g) \geq \frac{u_i(\resource)}{n}$, where the \emph{indivisibility ratio}~$\alpha_i$ for agent~$i$ is defined as $\alpha_i \coloneqq \frac{u_i(\indivisibleItems)}{u_i(\resource)}$.
\end{definition}

We can see from the above definition that the indivisibility ratio of an agent is smaller if she has a higher utility for the divisible goods.
This, in turn, implies that she is more likely to receive an allocation closer to proportionality.
One can also easily verify that PROP-$\alpha$ reduces to proportionality (resp., PROP1) if the resource consists of only divisible goods (resp., indivisible goods).
Furthermore, a PROP-$\alpha$ allocation can be efficiently computed, and a PROP-$\alpha$ and PO allocation always exists.

\begin{theorem}[\citet{LiLiLi24}]
Given any mixed-goods instance, a PROP-$\alpha$ allocation can be computed in polynomial time with polynomially many Robertson-Webb queries, and a PROP-$\alpha$ and PO allocation always exists via the maximum Nash welfare allocation.
\end{theorem}

\citet{LiLiLi24} also explored the \emph{tight} connection between \EFM (\Cref{def:EFM-goods}) and PROP-$\alpha$ (\Cref{def:PROPalpha}): \EFM $\implies$ PROP-$\alpha$.
Specifically, they showed that an \EFM allocation is PROP-$\alpha$, but for any $\epsilon > 0$, an \EFM allocation may not be PROP-$(1-\epsilon)\alpha$.
Here, PROP-$(1-\epsilon)\alpha$ is defined similarly to \Cref{def:PROPalpha}, except that $\alpha$-fraction of one good is replaced with $(1-\epsilon)\alpha$-fraction of one good.
We remark that although PROP-$\alpha$ is a weaker fairness notion than \EFM, it offers several advantages.
First, an allocation satisfying PROP-$\alpha$ can be efficiently found, while efficiently computing an \EFM allocation remains an open question (see \Cref{problem:bounded-EFM-alg}).
Second, PROP-$\alpha$ is compatible with PO, while it is an open question that whether \EFM and PO are compatible (see \Cref{problem:EFM-PO}).

\medskip

To conclude, the mixed-goods (or mixed-resources) model is rich and opens up new research directions that deserve further studies.
For instance, going beyond \EFM and MMS, can we define and study other meaningful fairness notions in the mixed-goods (or resources) model?
To this end, \citet{KawaseNiSu24} studied fair mixed-goods allocations whose utility vectors minimize a symmetric strictly convex function.
In a different direction, \citet{BeiLiLu23} further extended the mixed-goods model by letting agents have their own \emph{subjective divisibility} over the goods.
That is, some agents may find a good to be indivisible and get utilities only if they receive the \emph{whole} good, while other agents consider the same good as divisible and accumulate utilities in proportion to the fraction of the good they receive.

\section{Indivisible Goods with Subsidy}
\label{sec:subsidy}

In this section, we discuss how to allocate indivisible goods fairly through monetary compensation.
As money can be thought of as a homogeneous divisible good, this setting fits into the framework of mixed-goods setting studied in \Cref{sec:mixed-div-ind}.
The key difference in this section is that we consider money as a tool to achieve envy-freeness rather than an exogenously given resource to be divided fairly.

As envy-freeness---the quintessential notion of fairness in fair division---cannot be guaranteed when the goods are indivisible, many economists have attempted to circumvent this issue by introducing monetary compensation~\citep{Maskin87,Klijn00}.
However, earlier works in this line of research have mainly focused on the unit demand setting, wherein each agent is only interested in at most one good.
The setting of arbitrary number of goods under general additive valuations was considered only recently by \citet{HalpernSh19}.

Let us first discuss what it means to be fair in the presence of monetary compensations (also called subsidy payments).
We write $p = (p_1, p_2, \dots, p_n) \in \mathbb{R}^n_{\geq 0}$ as the vector of subsidy payments given to each agent, where $p_i$ denotes the subsidy payment given to agent~$i$.
The notion of envy-freeness with subsidy payment is defined as follows:

\begin{definition}
An allocation with payments $(\mathcal{\indivisibleItems}, p)$ is \emph{envy-free} if for any pair of agents~$i, j \in N$, $u_i(\indivisibleItems_i) + p_i \geq u_i(\indivisibleItems_j) + p_j$.
\end{definition}

In other words, an allocation with payments is envy-free if every agent prefers their own bundle plus payment to the bundle plus payment of any other agent.

It is important to note that not all allocations can be made envy-free by introducing payments.
For example, consider an instance with two agents~$1$ and~$2$, a single good~$g$, and $u_1(g) > u_2(g)$.
If the good is allocated to agent~$2$, then no subsidy payments $(p_1, p_2)$ exist so that the resulting allocation with payments is envy-free.
An allocation that can be made envy-free by introducing payments is called \emph{envy-freeable}.
\citet{HalpernSh19} showed the following characterization of envy-freeable allocations:

\begin{theorem}[\citet{HalpernSh19}]
\label{thm:HalpernSh19}
The following statements are equivalent:
\begin{enumerate}[label=(\roman*)]
\item The allocation~$\mathcal{\indivisibleItems}$ is envy-freeable.

\item The allocation~$\mathcal{\indivisibleItems}$ maximizes utilitarian welfare among all reassignments of the bundles, i.e., for every permutation $\sigma$ of the agents, $\sum_{i = 1}^n u_i(\indivisibleItems_i) \geq \sum_{i = 1}^n u_i(\indivisibleItems_{\sigma(i)})$.

\item The envy graph~$G_{\mathcal{\indivisibleItems}}$ contains no positive-weight directed cycle.\footnote{In \citep{HalpernSh19}, given an allocation~$\mathcal{\indivisibleItems}$, its envy graph~$G_{\mathcal{\indivisibleItems}}$ is the complete weighted directed graph in which for each pair of agents~$i, j \in N$, directed edge~$(i, j)$ has weight~$w(i, j) = u_i(\indivisibleItems_j) - u_i(\indivisibleItems_i)$.}
\end{enumerate}
\end{theorem}

An immediate consequence of \Cref{thm:HalpernSh19} is that any allocation can be made envy-freeable by reassigning the bundles.
Furthermore, \citet{HalpernSh19} showed that for a fixed envy-freeable allocation $\mathcal{\indivisibleItems}$, setting $p_i = \ell_{G_{\mathcal{\indivisibleItems}}}(i)$ not only makes $(\mathcal{\indivisibleItems}, p)$ envy-free but also minimizes the total subsidy required for doing so.
Here, $\ell_{G_{\mathcal{\indivisibleItems}}}(i)$ denotes the maximum weight of any path starting from node~$i$ in $G_{\mathcal{\indivisibleItems}}$.

Considering budgetary limitations of the mechanism designer, it is natural to study how much subsidy payment is required to guarantee envy-freeness.
\citet{HalpernSh19} conjectured that under additive valuations, subsidy of $n-1$ always suffices.\footnote{Each good is worth at most~$1$ for every agent.
This is achieved without loss of generality through a scaling argument.
Without scaling, the bound becomes $(n-1) \times \max_{i \in N, g \in \indivisibleItems} u_i(g)$.}
\citet{BrustleDiNa20} affirmatively settled this conjecture, where they showed an even stronger result:

\begin{theorem}[\citet{BrustleDiNa20}]
\label{thm:Subsidy}
For additive utilities, there exists a polynomial-time algorithm which outputs an envy-free allocation with subsidy $(\mathcal{\indivisibleItems}, p)$ such that:
\begin{enumerate}[label=(\roman*)]
\item Subsidy to each agent is at most one, i.e., $p_i \leq 1$.

\item Allocation~$\mathcal{\indivisibleItems}$ is \EFOne and balanced (i.e., $||O_i| - |O_j|| \leq 1$ for any~$i, j \in N$).
\end{enumerate}
\end{theorem}

Observe that \Cref{thm:Subsidy} implies the conjecture of \citeauthor{HalpernSh19}.
This is because if a subsidy payment eliminates envy, then these payments can be uniformly lowered while maintaining envy-freeness.
Hence, there is at least one agent who gets zero subsidy, which makes the total subsidy at most~$n-1$.
Furthermore, the bound of~$n-1$ on the subsidy required to guarantee the existence of envy-free allocations is tight.
To see this, consider an instance with a single good and $n$ agents who all value the good at~$1$.
For this instance, any envy-free allocation with subsidy must have a  total subsidy of at least~$n-1$.

The subsidy needed to guarantee envy-freeness is much less understood for valuation classes that are beyond additive.
\citet{BrustleDiNa20} showed that for monotone valuations, a total subsidy of~$2 (n-1)^2$ suffices to guarantee envy-free allocations.
Subsequently, \citet{KawaseMaSu24} improved this bound to $\frac{n^2-n-1}{2}$.\footnote{\citet{KawaseMaSu24}'s result works for doubly-monotonic utilities.}
As there are no lower bounds known beyond the aforementioned $n-1$ bound, this leads to a natural question.

\begin{problem}
For monotonic utilities, does there exist an envy-free allocation whose total subsidy is~$O(n^{2-\epsilon})$ for some $\epsilon > 0$?
\end{problem}

There has been progress made towards the above problem in restricted domains.
\citet{GokoIgKa24} showed that when the valuation functions are submodular functions with binary marginals (i.e., matroid-rank valuations), a total subsidy of~$n-1$ suffices.
Their mechanism additionally satisfies truthfulness.
In a subsequent work, \citet{BarmanKrNa22} showed that for general set valuations with binary marginals total subsidy payment of~$n-1$ suffices.

A natural and closely related direction is to study the optimization problem of computing an allocation using minimum total subsidy that achieves envy-freeness.
This problem is NP-hard since deciding whether an envy-free allocation exists for a given fair division instance is NP-hard.
The same argument shows that it is NP-hard to approximate the minimum subsidy to any multiplicative factor.
As a result, existing works have focused on additive approximation algorithms. \citet{CaragiannisIo21} showed that for constant number of agents, an $\varepsilon$ additive approximation algorithm can be computed in time polynomial in the number of goods and~$1 / \varepsilon$.
Furthermore, they showed that when the number of agents is part of the input, the problem is hard to approximate to within an additive factor of $c \sum_{i \in N} u_i(O)$ for some small constant~$c$.

Subsidy payments can also be studied for fairness notions other than envy-freeness. In a recent work, \citet{WuZhZh23} initiated the study of the minimum subsidy needed to guarantee the existence of a proportional allocation.\footnote{\citet{WuZhZh23}'s work mainly focused on chores; however, they also show that their subsidy bounds also hold for goods as well.} They showed that a total subsidy of $n/4$ suffices proportionality, in contrast to the $n-1$ subsidy needed for envy-freeness. In a subsequent work, \citet{WuZh24} strengthened the subsidy bounds needed to guarantee the existence of a weighted proportional allocation. It should be noted that there are multiple ways of defining share-based notions of fairness in the presence of subsidy, and they differ from each other in subtle ways. \citet{WuZhZh23} defined \emph{proportionality} as $u_i(O_i) + p_i \geq \frac{u_i(O)}{n}$ for each agent~$i \in N$. Here, the total subsidy is not included in the proportional share of an agent. Another possible way is to consider both the divisible good (total subsidy) and the indivisible good in the definition of proportional share, under which proportionality is defined as $u_i(O_i) + p_i \geq \frac{1}{n}(u_i(O) + \sum_{j \in N} p_j)$ for each agent~$i \in N$. In the latter definition of proportionality, it can be seen that subsidy of $n-1$ is  needed to guarantee the existence. Exploring other fairness notions (e.g., MMS and AnyPrice share~\citep{BabaioffEzFe21}) using subsidy payments is an intriguing direction for future research.

For a mechanism to utilize subsidy payments, it is necessary for the mechanism to possess sufficient funds to disburse such subsidies.
In many settings, however, the mechanism may not have access to adequate funds, making it difficult to implement. Such an issue can be circumvented if we allow for negative payments and additionally require $\sum_{i\in N} p_i=0$. These types of payments are referred to as \textit{transfer payments}. It can be seen that subsidy payments and transfer payments are interchangeable since whenever there is an envy-free allocation with subsidies, subtracting the average subsidy from each agent's individual payment results in payments that sum to zero and remains envy-free. \citet{NarayanSuVe21} studied whether transfer payments can be used to achieve both fairness and efficiency.\footnote{Transfer payments are better suited for studying welfare notions because they do not alter the social welfare of an allocation.} They showed that, for general monotone valuations, there exists an envy-free allocation with transfer payments whose Nash social welfare is at least $e^{-\frac{1}{e}}$-fraction of the optimal Nash social welfare. As for utilitarian social welfare, they give algorithms to compute an envy-free allocation with transfers that achieves a prescribed target welfare with a near-optimal bound on the amount of total transfer payments $\sum_{i\in N}|p_i|$ needed. In a related work, \citet{Aziz21} showed that transfer payments can be used to give an allocation that is both envy-free and equitable provided that the valuation function is supermodular.
He also studied various axiomatic properties of allocations that can be made both envy-free and equitable.

As seen from this section, by introducing a small amount of subsidy (or transfer) payments, one can achieve stronger fairness guarantees that are not possible otherwise in the indivisible items setting.
It is an interesting avenue of research to explore different settings for which subsidy payments can be helpful.
For instance, we may consider the indivisible items setting with externalities, where the value that an agent has for an allocation depends not only on their own bundle but also on the bundles allocated to everyone else.
Can subsidy payments be used to find fair allocations for problems with externalities?

\section{Conclusion}

In this survey, we have discussed several mixed fair division settings that generalize classical models in different ways, capture various realistic aspects of real-world scenarios, require non-trivial examinations of appropriate and attracting fairness concepts, and open up opportunities for a number of intriguing technical questions.
As we have seen in \Cref{sec:mixed-div-ind,sec:subsidy}, divisible resources to some extent help achieve stronger fairness properties.
In a similar vein, \Cref{sec:indivisible-items,sec:mixed-div-ind} demonstrate that approximate fairness can still be achieved with mixed types of resources.
However, simultaneously achieving approximate envy-freeness \emph{and} PO is a challenging problem in both mixed fair division settings, in contrast to, e.g., the classic setting with indivisible goods.

In addition to open questions outlined already, we present some other interesting directions below.
One direction is to allow practical allocation constraints; we refer interested readers to the recent survey of \citet{Suksompong21}.
Going beyond the context of dividing resources among agents, the idea of combining mixed types of resources has been investigated in a \emph{collective choice} context~\citep{LuPeAz24}, where all agents share a selected subset of the resources.
Extending the idea further to more general settings of allocating \emph{public} resources~\citep[see, e.g.,][on \emph{participatory budgeting}]{AzizSh21,ReyMa23}, or even to \emph{public decision making}~\citep{ConitzerFrSh17,SkowronGo22} is an interesting and practical direction.

\section*{Acknowledgments}

A preliminary version of this survey appeared as~\citep{LiuLuSu24}.
We would like to thank Ayumi Igarashi, Conrad Heilmann, Hadi Hosseini, Bo Li, Warut Suksompong, Rohit Vaish, Xiaowei Wu, and the anonymous reviewers for helpful comments and valuable feedback.

This work was partially supported by ARC Laureate Project FL200100204 on ``Trustworthy AI'', by the National Natural Science Foundation of China (Grant No.\ 62102117), by the Shenzhen Science and Technology Program (Grant Nos.\ RCBS20210609103900003 and GXWD20231129111306002), by the Guangdong Basic and Applied Basic Research Foundation (Grant No.\ 2023A1515011188), and by the CCF-Huawei Populus Grove Fund (Grant No.\ CCF-HuaweiLK2022005).

\bibliographystyle{plainnat}
\bibliography{bibliography}

\begin{thebibliography}{138}
\providecommand{\natexlab}[1]{#1}
\providecommand{\url}[1]{\texttt{#1}}
\expandafter\ifx\csname urlstyle\endcsname\relax
  \providecommand{\doi}[1]{doi: #1}\else
  \providecommand{\doi}{doi: \begingroup \urlstyle{rm}\Url}\fi

\bibitem[Akrami and Garg(2024)]{AkramiGa24}
Hannaneh Akrami and Jugal Garg.
\newblock Breaking the $3/4$ barrier for approximate maximin share.
\newblock In \emph{Proceedings of the 35th ACM-SIAM Symposium on Discrete
  Algorithms (SODA)}, pages 74--91, 2024.

\bibitem[Akrami et~al.(2023{\natexlab{a}})Akrami, Alon, Chaudhury, Garg,
  Mehlhorn, and Mehta]{AkramiAlCh23}
Hannaneh Akrami, Noga Alon, Bhaskar~Ray Chaudhury, Jugal Garg, Kurt Mehlhorn,
  and Ruta Mehta.
\newblock {EFX}: {A} simpler approach and an (almost) optimal guarantee via
  rainbow cycle number.
\newblock In \emph{Proceedings of the 24th ACM Conference on Economics and
  Computation (EC)}, page~61, 2023{\natexlab{a}}.

\bibitem[Akrami et~al.(2023{\natexlab{b}})Akrami, Mehlhorn, Seddighin, and
  Shahkarami]{AkramiMeSe23}
Hannaneh Akrami, Kurt Mehlhorn, Masoud Seddighin, and Golnoosh Shahkarami.
\newblock Randomized and deterministic maximin-share approximations for
  fractionally subadditive valuations.
\newblock In \emph{Proceedings of the 37th Annual Conference on Neural
  Information Processing Systems (NeurIPS)}, pages 58821--58832,
  2023{\natexlab{b}}.

\bibitem[Akrami et~al.(2024)Akrami, Garg, Sharma, and Taki]{AkramiGaSh24}
Hannaneh Akrami, Jugal Garg, Eklavya Sharma, and Setareh Taki.
\newblock Improving approximation guarantees for maximin share.
\newblock In \emph{Proceedings of the 25th ACM Conference on Economics and
  Computation (EC)}, 2024.
\newblock Forthcoming.

\bibitem[Aleksandrov and Walsh(2020)]{AleksandrovWa20-KI}
Martin Aleksandrov and Toby Walsh.
\newblock Two algorithms for additive and fair division of mixed manna.
\newblock In \emph{Proceedings of the 43rd German Conference on Artificial
  Intelligence (KI)}, pages 3--17, 2020.

\bibitem[Alon(1987)]{Alon87}
Noga Alon.
\newblock Splitting necklaces.
\newblock \emph{Advances in Mathematics}, 63\penalty0 (3):\penalty0 247--253,
  1987.

\bibitem[Amanatidis et~al.(2017)Amanatidis, Birmpas, Christodoulou, and
  Markakis]{AmanatidisBiCh17}
Georgios Amanatidis, Georgios Birmpas, George Christodoulou, and Evangelos
  Markakis.
\newblock Truthful allocation mechanisms without payments: Characterization and
  implications on fairness.
\newblock In \emph{Proceedings of the 18th ACM Conference on Economics and
  Computation (EC)}, pages 545--562, 2017.

\bibitem[Amanatidis et~al.(2023)Amanatidis, Aziz, Birmpas, Filos-Ratsikas, Li,
  Moulin, Voudouris, and Wu]{AmanatidisAzBi23}
Georgios Amanatidis, Haris Aziz, Georgios Birmpas, Aris Filos-Ratsikas, Bo~Li,
  Herv\'{e} Moulin, Alexandros~A. Voudouris, and Xiaowei Wu.
\newblock Fair division of indivisible goods: Recent progress and open
  questions.
\newblock \emph{Artificial Intelligence}, 322:\penalty0 103965, 2023.

\bibitem[Arrow and Debreu(1954)]{ArrowDe54}
Kenneth~J. Arrow and Gerard Debreu.
\newblock Existence of an equilibrium for a competitive economy.
\newblock \emph{Econometrica}, 22\penalty0 (3):\penalty0 265--290, 1954.

\bibitem[Aziz(2020)]{Aziz20}
Haris Aziz.
\newblock Developments in multi-agent fair allocation.
\newblock In \emph{Proceedings of the 34th AAAI Conference on Artificial
  Intelligence (AAAI)}, pages 13563--13568, 2020.

\bibitem[Aziz(2021)]{Aziz21}
Haris Aziz.
\newblock Achieving envy-freeness and equitability with monetary transfers.
\newblock In \emph{Proceedings of the 34th AAAI Conference on Artificial
  Intelligence (AAAI)}, pages 5102--5109, 2021.

\bibitem[Aziz and Rey(2020)]{AzizRe20}
Haris Aziz and Simon Rey.
\newblock Almost group envy-free allocation of indivisible goods and chores.
\newblock In \emph{Proceedings of the 29th International Joint Conference on
  Artificial Intelligence (IJCAI)}, pages 39--45, 2020.

\bibitem[Aziz and Shah(2021)]{AzizSh21}
Haris Aziz and Nisarg Shah.
\newblock Participatory budgeting: Models and approaches.
\newblock In Tam\'{a}s Rudas and G\'{a}bor P\'{e}li, editors, \emph{Pathways
  Between Social Science and Computational Social Science: Theories, Methods,
  and Interpretations}, pages 215--236. Springer International Publishing,
  2021.

\bibitem[Aziz et~al.(2019)Aziz, Bir\'{o}, Lang, Lesca, and Monnot]{AzizBiLa19}
Haris Aziz, P\'{e}ter Bir\'{o}, J\'{e}r\^{o}me Lang, Julien Lesca, and
  J\'{e}r\^{o}me Monnot.
\newblock Efficient reallocation under additive and responsive preferences.
\newblock \emph{Theoretical Computer Science}, 790:\penalty0 1--15, 2019.

\bibitem[Aziz et~al.(2020)Aziz, Moulin, and Sandomirskiy]{AzizMoSa20}
Haris Aziz, Herv\'{e} Moulin, and Fedor Sandomirskiy.
\newblock A polynomial-time algorithm for computing a {P}areto optimal and
  almost proportional allocation.
\newblock \emph{Operations Research Letters}, 48\penalty0 (5):\penalty0
  573--578, 2020.

\bibitem[Aziz et~al.(2022)Aziz, Caragiannis, Igarashi, and Walsh]{AzizCaIg22}
Haris Aziz, Ioannis Caragiannis, Ayumi Igarashi, and Toby Walsh.
\newblock Fair allocation of indivisible goods and chores.
\newblock \emph{Autonomous Agents and Multi-Agent Systems}, 36\penalty0
  (1):\penalty0 3:1--3:21, 2022.

\bibitem[Aziz et~al.(2023{\natexlab{a}})Aziz, Freeman, Shah, and
  Vaish]{AzizFrSh23}
Haris Aziz, Rupert Freeman, Nisarg Shah, and Rohit Vaish.
\newblock Best of both worlds: Ex-ante and ex-post fairness in resource
  allocation.
\newblock \emph{Operations Research}, 2023{\natexlab{a}}.
\newblock Forthcoming.

\bibitem[Aziz et~al.(2023{\natexlab{b}})Aziz, Ganguly, and Micha]{AzizGaMi23}
Haris Aziz, Aditya Ganguly, and Evi Micha.
\newblock Best of both worlds fairness under entitlements.
\newblock In \emph{Proceedings of the 22nd International Conference on
  Autonomous Agents and Multiagent Systems (AAMAS)}, pages 941--948,
  2023{\natexlab{b}}.

\bibitem[Aziz et~al.(2023{\natexlab{c}})Aziz, Huang, Mattei, and
  Segal-Halevi]{AzizHuMa23}
Haris Aziz, Xin Huang, Nicholas Mattei, and Erel Segal-Halevi.
\newblock Computing welfare-maximizing fair allocations of indivisible goods.
\newblock \emph{European Journal of Operational Research}, 307\penalty0
  (2):\penalty0 773--784, 2023{\natexlab{c}}.

\bibitem[Aziz et~al.(2023{\natexlab{d}})Aziz, Lindsay, Ritossa, and
  Suzuki]{AzizLiRi23}
Haris Aziz, Jeremy Lindsay, Angus Ritossa, and Mashbat Suzuki.
\newblock Fair allocation of two types of chores.
\newblock In \emph{Proceedings of the 22nd International Conference on
  Autonomous Agents and Multiagent Systems (AAMAS)}, pages 143--151,
  2023{\natexlab{d}}.

\bibitem[Aziz et~al.(2023{\natexlab{e}})Aziz, Lu, Suzuki, Vollen, and
  Walsh]{AzizLuSu23}
Haris Aziz, Xinhang Lu, Mashbat Suzuki, Jeremy Vollen, and Toby Walsh.
\newblock Best-of-both-worlds fairness in committee voting.
\newblock In \emph{Proceedings of the 19th Conference on Web and Internet
  Economics (WINE)}, page 676, 2023{\natexlab{e}}.

\bibitem[Aziz et~al.(2023{\natexlab{f}})Aziz, Suksompong, Sun, and
  Walsh]{AzizSuSu23}
Haris Aziz, Warut Suksompong, Zhaohong Sun, and Toby Walsh.
\newblock Fairness concepts for indivisible items with externalities.
\newblock In \emph{Proceedings of the 37th AAAI Conference on Artificial
  Intelligence (AAAI)}, pages 5472--5480, 2023{\natexlab{f}}.

\bibitem[Aziz et~al.(2024{\natexlab{a}})Aziz, Li, Moulin, Wu, and
  Zhu]{AzizLiMo24}
Haris Aziz, Bo~Li, Herv\'{e} Moulin, Xiaowei Wu, and Xinran Zhu.
\newblock Almost proportional allocations of indivisible chores: Computation,
  approximation and efficiency.
\newblock \emph{Artificial Intelligence}, 331:\penalty0 104118,
  2024{\natexlab{a}}.

\bibitem[Aziz et~al.(2024{\natexlab{b}})Aziz, Lu, Suzuki, Vollen, and
  Walsh]{AzizLuSu24}
Haris Aziz, Xinhang Lu, Mashbat Suzuki, Jeremy Vollen, and Toby Walsh.
\newblock Fair lotteries for participatory budgeting.
\newblock In \emph{Proceedings of the 38th AAAI Conference on Artificial
  Intelligence (AAAI)}, pages 9469--9476, 2024{\natexlab{b}}.

\bibitem[Babaioff et~al.(2021)Babaioff, Ezra, and Feige]{BabaioffEzFe21}
Moshe Babaioff, Tomer Ezra, and Uriel Feige.
\newblock Fair and truthful mechanisms for dichotomous valuations.
\newblock In \emph{Proceedings of the 35th AAAI Conference on Artificial
  Intelligence (AAAI)}, pages 5119--5126, 2021.

\bibitem[Babaioff et~al.(2022)Babaioff, Ezra, and Feige]{BabaioffEzFe22}
Moshe Babaioff, Tomer Ezra, and Uriel Feige.
\newblock On best-of-both-worlds fair-share allocations.
\newblock In \emph{Proceedings of the 18th Conference on Web and Internet
  Economics (WINE)}, pages 237--255, 2022.

\bibitem[Barman and Verma(2022)]{BarmanVe22}
Siddharth Barman and Paritosh Verma.
\newblock Truthful and fair mechanisms for matroid-rank valuations.
\newblock In \emph{Proceedings of the 36th AAAI Conference on Artificial
  Intelligence (AAAI)}, pages 4801--4808, 2022.

\bibitem[Barman et~al.(2018)Barman, Krishnamurthy, and Vaish]{BarmanKrVa18}
Siddharth Barman, Sanath~Kumar Krishnamurthy, and Rohit Vaish.
\newblock Finding fair and efficient allocations.
\newblock In \emph{Proceedings of the 19th ACM Conference on Economics and
  Computation (EC)}, pages 557--574, 2018.

\bibitem[Barman et~al.(2020)Barman, Bhaskar, and Shah]{BarmanBhSh20}
Siddharth Barman, Umang Bhaskar, and Nisarg Shah.
\newblock Optimal bounds on the price of fairness for indivisible goods.
\newblock In \emph{Proceedings of the 16th International Conference on Web and
  Internet Economics (WINE)}, pages 356--369, 2020.

\bibitem[Barman et~al.(2022)Barman, Krishna, Narahari, and
  Sadhukhan]{BarmanKrNa22}
Siddharth Barman, Anand Krishna, Yadati Narahari, and Soumyarup Sadhukhan.
\newblock Achieving envy-freeness with limited subsidies under dichotomous
  valuations.
\newblock In \emph{Proceedings of the 31st International Joint Conference on
  Artificial Intelligence (IJCAI)}, pages 60--66, 2022.

\bibitem[Bei et~al.(2012)Bei, Chen, Hua, Tao, and Yang]{BeiChHu12}
Xiaohui Bei, Ning Chen, Xia Hua, Biaoshuai Tao, and Endong Yang.
\newblock Optimal proportional cake cutting with connected pieces.
\newblock In \emph{Proceedings of the 26th AAAI Conference on Artificial
  Intelligence (AAAI)}, pages 1263--1269, 2012.

\bibitem[Bei et~al.(2020)Bei, Huzhang, and Suksompong]{BeiHuSu20}
Xiaohui Bei, Guangda Huzhang, and Warut Suksompong.
\newblock Truthful fair division without free disposal.
\newblock \emph{Social Choice and Welfare}, 55:\penalty0 523--545, 2020.

\bibitem[Bei et~al.(2021{\natexlab{a}})Bei, Li, Liu, Liu, and Lu]{BeiLiLi21}
Xiaohui Bei, Zihao Li, Jinyan Liu, Shengxin Liu, and Xinhang Lu.
\newblock Fair division of mixed divisible and indivisible goods.
\newblock \emph{Artificial Intelligence}, 293:\penalty0 103436,
  2021{\natexlab{a}}.

\bibitem[Bei et~al.(2021{\natexlab{b}})Bei, Liu, Lu, and Wang]{BeiLiLu21}
Xiaohui Bei, Shengxin Liu, Xinhang Lu, and Hongao Wang.
\newblock Maximin fairness with mixed divisible and indivisible goods.
\newblock \emph{Autonomous Agents and Multi-Agent Systems}, 35\penalty0
  (2):\penalty0 34:1--34:21, 2021{\natexlab{b}}.

\bibitem[Bei et~al.(2021{\natexlab{c}})Bei, Lu, Manurangsi, and
  Suksompong]{BeiLuMa21}
Xiaohui Bei, Xinhang Lu, Pasin Manurangsi, and Warut Suksompong.
\newblock The price of fairness for indivisible goods.
\newblock \emph{Theory of Computing Systems}, 65\penalty0 (7):\penalty0
  1069--1093, 2021{\natexlab{c}}.

\bibitem[Bei et~al.(2023)Bei, Liu, and Lu]{BeiLiLu23}
Xiaohui Bei, Shengxin Liu, and Xinhang Lu.
\newblock Fair division with subjective divisibility.
\newblock In \emph{Proceedings of the 19th Conference on Web and Internet
  Economics (WINE)}, page 677, 2023.

\bibitem[Bei et~al.(2024)Bei, Lu, and Suksompong]{BeiLuSu24}
Xiaohui Bei, Xinhang Lu, and Warut Suksompong.
\newblock Truthful cake sharing.
\newblock \emph{Social Choice and Welfare}, 2024.
\newblock Forthcoming.

\bibitem[B\'{e}rczi et~al.(2020)B\'{e}rczi, B\'erczi-Kov\'acs, Boros, Gedefa,
  Kamiyama, Kavitha, Kobayashi, and Makino]{BercziBeBo20}
Krist\'{o}f B\'{e}rczi, Erika~R. B\'erczi-Kov\'acs, Endre Boros, Fekadu~Tolessa
  Gedefa, Naoyuki Kamiyama, Telikepalli Kavitha, Yusuke Kobayashi, and Kazuhisa
  Makino.
\newblock Envy-free relaxations for goods, chores, and mixed items.
\newblock \emph{CoRR}, abs/2006.04428, 2020.

\bibitem[Berliant et~al.(1992)Berliant, Thomson, and Dunz]{BerliantThDu92}
Marcus Berliant, William Thomson, and Karl Dunz.
\newblock On the fair division of a heterogeneous commodity.
\newblock \emph{Journal of Mathematical Economics}, 21\penalty0 (3):\penalty0
  201--216, 1992.

\bibitem[Bertsimas et~al.(2011)Bertsimas, Farias, and
  Trichakis]{BertsimasFaTr11}
Dimitris Bertsimas, Vivek~F. Farias, and Nikolaos Trichakis.
\newblock The price of fairness.
\newblock \emph{Operations Research}, 59\penalty0 (1):\penalty0 17--31, 2011.

\bibitem[Bhaskar et~al.(2021)Bhaskar, Sricharan, and Vaish]{BhaskarSrVa21}
Umang Bhaskar, A.~R. Sricharan, and Rohit Vaish.
\newblock On approximate envy-freeness for indivisible chores and mixed
  resources.
\newblock In \emph{Proceedings of the 24th International Conference on
  Approximation Algorithms for Combinatorial Optimization Problems (APPROX)},
  pages 1:1--1:23, 2021.

\bibitem[Bogomolnaia and Moulin(2004)]{BogomolnaiaMo04}
Anna Bogomolnaia and Herv\'{e} Moulin.
\newblock Random matching under dichotomous preferences.
\newblock \emph{Econometrica}, 72\penalty0 (1):\penalty0 257--279, 2004.

\bibitem[Bogomolnaia et~al.(2017)Bogomolnaia, Moulin, Sandomirskiy, and
  Yanovskaya]{BogomolnaiaMoSa17}
Anna Bogomolnaia, Herv\'{e} Moulin, Fedor Sandomirskiy, and Elena Yanovskaya.
\newblock Competitive division of a mixed manna.
\newblock \emph{Econometrica}, 85\penalty0 (6):\penalty0 1847--1871, 2017.

\bibitem[Brams and Taylor(1996)]{BramsTa96}
Steven~J. Brams and Alan~D. Taylor.
\newblock \emph{Fair Division: From Cake-Cutting to Dispute Resolution}.
\newblock Cambridge University Press, 1996.

\bibitem[Brandl et~al.(2021)Brandl, Brandt, Peters, and Stricker]{BrandlBrPe21}
Florian Brandl, Felix Brandt, Dominik Peters, and Christian Stricker.
\newblock Distribution rules under dichotomous preferences: Two out of three
  ain't bad.
\newblock In \emph{Proceedings of the 22nd ACM Conference on Economics and
  Computation (EC)}, pages 158--179, 2021.

\bibitem[Brandt et~al.(2016)Brandt, Conitzer, Endriss, Lang, and
  Procaccia]{BrandtCoEn16}
Felix Brandt, Vincent Conitzer, Ulle Endriss, J\'{e}r\^{o}me Lang, and Ariel~D.
  Procaccia, editors.
\newblock \emph{Handbook of Computational Social Choice}.
\newblock Cambridge University Press, 2016.

\bibitem[Br\^{a}nzei et~al.(2013)Br\^{a}nzei, Procaccia, and
  Zhang]{BranzeiPrZh13}
Simina Br\^{a}nzei, Ariel~D. Procaccia, and Jie Zhang.
\newblock Externalities in cake cutting.
\newblock In \emph{Proceedings of the 23rd International Joint Conference on
  Artificial Intelligence (IJCAI)}, pages 55--61, 2013.

\bibitem[Brustle et~al.(2020)Brustle, Dippel, Narayan, Suzuki, and
  Vetta]{BrustleDiNa20}
Johannes Brustle, Jack Dippel, Vishnu~V. Narayan, Mashbat Suzuki, and Adrian
  Vetta.
\newblock One dollar each eliminates envy.
\newblock In \emph{Proceedings of the 21st ACM Conference on Economics and
  Computation (EC)}, pages 23--39, 2020.

\bibitem[Bu et~al.(2023{\natexlab{a}})Bu, Li, Liu, Song, and Tao]{BuLiLi23}
Xiaolin Bu, Zihao Li, Shengxin Liu, Jiaxin Song, and Biaoshuai Tao.
\newblock On the complexity of maximizing social welfare within fair
  allocations of indivisible goods.
\newblock \emph{CoRR}, abs/2205.14296, 2023{\natexlab{a}}.

\bibitem[Bu et~al.(2023{\natexlab{b}})Bu, Song, and Tao]{BuSoTa23}
Xiaolin Bu, Jiaxin Song, and Biaoshuai Tao.
\newblock On existence of truthful fair cake cutting mechanisms.
\newblock \emph{Artificial Intelligence}, 319:\penalty0 103904,
  2023{\natexlab{b}}.

\bibitem[Budish(2011)]{Budish11}
Eric Budish.
\newblock The combinatorial assignment problem: Approximate competitive
  equilibrium from equal incomes.
\newblock \emph{Journal of Political Economy}, 119\penalty0 (6):\penalty0
  1061--1103, 2011.

\bibitem[Budish et~al.(2017)Budish, Cachon, Kessler, and Othman]{BudishCaKe17}
Eric Budish, G\'{e}rard~P. Cachon, Judd~B. Kessler, and Abraham Othman.
\newblock Course {M}atch: {A} large-scale implementation of approximate
  competitive equilibrium from equal incomes for combinatorial allocation.
\newblock \emph{Operations Research}, 65\penalty0 (2):\penalty0 314--336, 2017.

\bibitem[Caragiannis and Ioannidis(2021)]{CaragiannisIo21}
Ioannis Caragiannis and Stavros Ioannidis.
\newblock Computing envy-freeable allocations with limited subsidies.
\newblock In \emph{Proceedings of the 17th International Conference on Web and
  Internet Economics (WINE)}, pages 522--539, 2021.

\bibitem[Caragiannis and Narang(2024)]{CaragiannisNa24}
Ioannis Caragiannis and Shivika Narang.
\newblock Repeatedly matching items to agents fairly and efficiently.
\newblock \emph{Theoretical Computer Science}, 981:\penalty0 114246, 2024.

\bibitem[Caragiannis et~al.(2012)Caragiannis, Kaklamanis, Kanellopoulos, and
  Kyropoulou]{CaragiannisKaKa12}
Ioannis Caragiannis, Christos Kaklamanis, Panagiotis Kanellopoulos, and Maria
  Kyropoulou.
\newblock The efficiency of fair division.
\newblock \emph{Theory of Computing Systems}, 50\penalty0 (4):\penalty0
  589--610, 2012.

\bibitem[Caragiannis et~al.(2019)Caragiannis, Kurokawa, Moulin, Procaccia,
  Shah, and Wang]{CaragiannisKuMo19}
Ioannis Caragiannis, David Kurokawa, Herv\'{e} Moulin, Ariel~D. Procaccia,
  Nisarg Shah, and Junxing Wang.
\newblock The unreasonable fairness of maximum {N}ash welfare.
\newblock \emph{ACM Transactions on Economics and Computation}, 7\penalty0
  (3):\penalty0 12:1--12:32, 2019.

\bibitem[Chaudhury et~al.(2023)Chaudhury, Garg, McGlaughlin, and
  Mehta]{ChaudhuryGaMc23}
Bhaskar~Ray Chaudhury, Jugal Garg, Peter McGlaughlin, and Ruta Mehta.
\newblock A complementary pivot algorithm for competitive allocation of a mixed
  manna.
\newblock \emph{Mathematics of Operations Research}, 48\penalty0 (3):\penalty0
  1630--1656, 2023.

\bibitem[Chaudhury et~al.(2024)Chaudhury, Garg, and Mehlhorn]{ChaudhuryGaMe24}
Bhaskar~Ray Chaudhury, Jugal Garg, and Kurt Mehlhorn.
\newblock {EFX} exists for three agents.
\newblock \emph{Journal of the ACM}, 71\penalty0 (1):\penalty0 4:1--4:27, 2024.

\bibitem[Chen et~al.(2013)Chen, Lai, Parkes, and Procaccia]{ChenLaPa13}
Yiling Chen, John~K. Lai, David~C. Parkes, and Ariel~D. Procaccia.
\newblock Truth, justice, and cake cutting.
\newblock \emph{Games and Economic Behavior}, 77\penalty0 (1):\penalty0
  284--297, 2013.

\bibitem[Christoforidis and Santorinaios(2024)]{ChristoforidisSa24}
Vasilis Christoforidis and Christodoulos Santorinaios.
\newblock On the pursuit of {EFX} for chores: Non-existence and approximations.
\newblock In \emph{Proceedings of the 33rd International Joint Conference on
  Artificial Intelligence (IJCAI)}, pages 2713--2721, 2024.

\bibitem[Cohler et~al.(2011)Cohler, Lai, Parkes, and Procaccia]{CohlerLaPa11}
Yuga~J. Cohler, John~K. Lai, David~C. Parkes, and Ariel~D. Procaccia.
\newblock Optimal envy-free cake cutting.
\newblock In \emph{Proceedings of the 25th AAAI Conference on Artificial
  Intelligence (AAAI)}, pages 626--631, 2011.

\bibitem[Conitzer et~al.(2017)Conitzer, Freeman, and Shah]{ConitzerFrSh17}
Vincent Conitzer, Rupert Freeman, and Nisarg Shah.
\newblock Fair public decision making.
\newblock In \emph{Proceedings of the 18th ACM Conference on Economics and
  Computation (EC)}, pages 629--646, 2017.

\bibitem[Cousins et~al.(2023)Cousins, Viswanathan, and Zick]{CousinsViZi23}
Cyrus Cousins, Vignesh Viswanathan, and Yair Zick.
\newblock The good, the bad and the submodular: Fairly allocating mixed manna
  under order-neutral submodular preferences.
\newblock In \emph{Proceedings of the 19th Conference on Web and Internet
  Economics (WINE)}, pages 207--224, 2023.

\bibitem[de~Keijzer et~al.(2009)de~Keijzer, Bouveret, Klos, and
  Zhang]{deKeijzerBoKl09}
Bart de~Keijzer, Sylvain Bouveret, Tomas Klos, and Yingqian Zhang.
\newblock On the complexity of efficiency and envy-freeness in fair division of
  indivisible goods with additive preferences.
\newblock In \emph{Proceedings of the 1st International Conference on
  Algorithmic Decision Theory (ADT)}, pages 98--110, 2009.

\bibitem[Ebadian et~al.(2022)Ebadian, Peters, and Shah]{EbadianPeSh22}
Soroush Ebadian, Dominik Peters, and Nisarg Shah.
\newblock How to fairly allocate easy and difficult chores.
\newblock In \emph{Proceedings of the 21st International Conference on
  Autonomous Agents and Multiagent Systems (AAMAS)}, pages 372--380, 2022.

\bibitem[Endriss(2017)]{Endriss17}
Ulle Endriss, editor.
\newblock \emph{Trends in Computational Social Choice}.
\newblock AI Access, 2017.

\bibitem[Feldman et~al.(2024)Feldman, Mauras, Narayan, and
  Ponitka]{FeldmanMaNa24}
Michal Feldman, Simon Mauras, Vishnu~V. Narayan, and Tomasz Ponitka.
\newblock Breaking the envy cycle: Best-of-both-worlds guarantees for
  subadditive valuations.
\newblock In \emph{Proceedings of the 25th ACM Conference on Economics and
  Computation (EC)}, 2024.
\newblock Forthcoming.

\bibitem[Foley(1967)]{Foley67}
Duncan~Karl Foley.
\newblock Resource allocation and the public sector.
\newblock \emph{Yale Economics Essays}, 7\penalty0 (1):\penalty0 45--98, 1967.

\bibitem[Freeman and Schmidt-Kraepelin(2024)]{FreemanSc24}
Rupert Freeman and Ulrike Schmidt-Kraepelin.
\newblock Project-fair and truthful mechanisms for budget aggregation.
\newblock In \emph{Proceedings of the 38th AAAI Conference on Artificial
  Intelligence (AAAI)}, pages 9704--9712, 2024.

\bibitem[Freeman et~al.(2021{\natexlab{a}})Freeman, Micha, and
  Shah]{FreemanMiSh21}
Rupert Freeman, Evi Micha, and Nisarg Shah.
\newblock Two-sided matching meets fair division.
\newblock In \emph{Proceedings of the 30th International Joint Conference on
  Artificial Intelligence (IJCAI)}, pages 203--209, 2021{\natexlab{a}}.

\bibitem[Freeman et~al.(2021{\natexlab{b}})Freeman, Pennock, Peters, and
  Wortman~Vaughan]{FreemanPePe21}
Rupert Freeman, David~M. Pennock, Dominik Peters, and Jennifer Wortman~Vaughan.
\newblock Truthful aggregation of budget proposals.
\newblock \emph{Journal of Economic Theory}, 193:\penalty0 105234,
  2021{\natexlab{b}}.

\bibitem[Friedman et~al.(2019)Friedman, Gkatzelis, Psomas, and
  Shenker]{FriedmanGkPs19}
Eric~J. Friedman, Vasilis Gkatzelis, Christos-Alexandros Psomas, and Scott
  Shenker.
\newblock Fair and efficient memory sharing: Confronting free riders.
\newblock In \emph{Proceedings of the 33rd AAAI Conference on Artificial
  Intelligence (AAAI)}, pages 1965--1972, 2019.

\bibitem[Garg and McGlaughlin(2020)]{GargMc20}
Jugal Garg and Peter McGlaughlin.
\newblock Computing competitive equilibria with mixed manna.
\newblock In \emph{Proceedings of the 19th International Conference on
  Autonomous Agents and Multiagent Systems (AAMAS)}, pages 420--428, 2020.

\bibitem[Garg et~al.(2021)Garg, Hoefer, McGlaughlin, and
  Schmalhofer]{GargHoMc21}
Jugal Garg, Martin Hoefer, Peter McGlaughlin, and Marco Schmalhofer.
\newblock When dividing mixed manna is easier than dividing goods: Competitive
  equilibria with a constant number of chores.
\newblock In \emph{Proceedings of the 14th International Symposium on
  Algorithmic Game Theory (SAGT)}, pages 329--344, 2021.

\bibitem[Garg et~al.(2022)Garg, Murhekar, and Qin]{GargMuQi22}
Jugal Garg, Aniket Murhekar, and John Qin.
\newblock Fair and efficient allocations of chores under bivalued preferences.
\newblock In \emph{Proceedings of the 36th AAAI Conference on Artificial
  Intelligence (AAAI)}, pages 5043--5050, 2022.

\bibitem[Garg et~al.(2023)Garg, Murhekar, and Qin]{GargMuQi23}
Jugal Garg, Aniket Murhekar, and John Qin.
\newblock New algorithms for the fair and efficient allocation of indivisible
  chores.
\newblock In \emph{Proceedings of the 32nd International Joint Conference on
  Artificial Intelligence (IJCAI)}, pages 2710--2718, 2023.

\bibitem[Goko et~al.(2024)Goko, Igarashi, Kawase, Makino, Sumita, Tamura,
  Yokoi, and Yokoo]{GokoIgKa24}
Hiromichi Goko, Ayumi Igarashi, Yasushi Kawase, Kazuhisa Makino, Hanna Sumita,
  Akihisa Tamura, Yu~Yokoi, and Makoto Yokoo.
\newblock A fair and truthful mechanism with limited subsidy.
\newblock \emph{Games and Economic Behavior}, 144:\penalty0 49--70, 2024.

\bibitem[Goldman and Procaccia(2015)]{GoldmanPr15}
Jonathan Goldman and Ariel~D. Procaccia.
\newblock Spliddit: Unleashing fair division algorithms.
\newblock \emph{SIGecom Exchanges}, 13\penalty0 (2):\penalty0 41--46, 2015.

\bibitem[Guo et~al.(2023)Guo, Li, and Deng]{GuoLiDe23}
Hao Guo, Weidong Li, and Bin Deng.
\newblock A survey on fair allocation of chores.
\newblock \emph{Mathematics}, 11\penalty0 (16):\penalty0 3616, 2023.

\bibitem[Halpern and Shah(2019)]{HalpernSh19}
Daniel Halpern and Nisarg Shah.
\newblock Fair division with subsidy.
\newblock In \emph{Proceedings of the 12th International Symposium on
  Algorithmic Game Theory (SAGT)}, pages 374--389, 2019.

\bibitem[Halpern et~al.(2020)Halpern, Procaccia, Psomas, and
  Shah]{HalpernPrPs20}
Daniel Halpern, Ariel~D. Procaccia, Alexandros Psomas, and Nisarg Shah.
\newblock Fair division with binary valuations: One rule to rule them all.
\newblock In \emph{Proceedings of the 16th International Conference on Web and
  Internet Economics (WINE)}, pages 370--383, 2020.

\bibitem[Han and Suksompong(2024)]{HanSu24}
Jiatong Han and Warut Suksompong.
\newblock Fast \& fair: A collaborative platform for fair division
  applications.
\newblock In \emph{Proceedings of the 38th AAAI Conference on Artificial
  Intelligence (AAAI)}, pages 23796--23798, 2024.
\newblock Demonstration Track.

\bibitem[Heilmann and Wintein(2021)]{HeilmannWi21}
Conrad Heilmann and Stefan Wintein.
\newblock No envy: {J}an {T}inbergen on fairness.
\newblock \emph{Erasmus Journal for Philosophy and Economics}, 14\penalty0
  (1):\penalty0 222--245, 2021.

\bibitem[Hoefer et~al.(2023)Hoefer, Schmalhofer, and Varricchio]{HoeferScVa23}
Martin Hoefer, Marco Schmalhofer, and Giovanna Varricchio.
\newblock Best of both worlds: Agents with entitlements.
\newblock In \emph{Proceedings of the 22nd International Conference on
  Autonomous Agents and Multiagent Systems (AAMAS)}, pages 564--572, 2023.

\bibitem[Hosseini et~al.(2023{\natexlab{a}})Hosseini, Mammadov, and
  W\k{a}s]{HosseiniMaWa23}
Hadi Hosseini, Aghaheybat Mammadov, and Tomasz W\k{a}s.
\newblock Fairly allocating goods and (terrible) chores.
\newblock In \emph{Proceedings of the 32nd International Joint Conference on
  Artificial Intelligence (IJCAI)}, pages 2738--2746, 2023{\natexlab{a}}.

\bibitem[Hosseini et~al.(2023{\natexlab{b}})Hosseini, Sikdar, Vaish, and
  Xia]{HosseiniSiVa23}
Hadi Hosseini, Sujoy Sikdar, Rohit Vaish, and Lirong Xia.
\newblock Fairly dividing mixtures of goods and chores under lexicographic
  preferences.
\newblock In \emph{Proceedings of the 22nd International Conference on
  Autonomous Agents and Multiagent Systems (AAMAS)}, pages 152--160,
  2023{\natexlab{b}}.

\bibitem[Huang and Segal-Halevi(2023)]{HuangSe23}
Xin Huang and Erel Segal-Halevi.
\newblock A reduction from chores allocation to job scheduling.
\newblock In \emph{Proceedings of the 24th ACM Conference on Economics and
  Computation (EC)}, page 908, 2023.

\bibitem[Igarashi and Yokoyama(2023)]{IgarashiYo23}
Ayumi Igarashi and Tomohiko Yokoyama.
\newblock Kajibuntan: {A} house chore division app.
\newblock In \emph{Proceedings of the 37th AAAI Conference on Artificial
  Intelligence (AAAI)}, pages 16449--16451, 2023.
\newblock Demo Track.

\bibitem[Igarashi et~al.(2023)Igarashi, Kawase, Suksompong, and
  Sumita]{IgarashiKaSu23}
Ayumi Igarashi, Yasushi Kawase, Warut Suksompong, and Hanna Sumita.
\newblock Fair division with two-sided preferences.
\newblock In \emph{Proceedings of the 32nd International Joint Conference on
  Artificial Intelligence (IJCAI)}, pages 2756--2764, 2023.

\bibitem[Jansen et~al.(2020)Jansen, Klein, and Verschae]{JansenKlVe20}
Klaus Jansen, Kim-Manuel Klein, and Jos\'{e} Verschae.
\newblock Closing the gap for makespan scheduling via sparsification
  techniques.
\newblock \emph{Mathematics of Operations Research}, 45\penalty0 (4):\penalty0
  1371--1392, 2020.

\bibitem[Kawase et~al.(2024{\natexlab{a}})Kawase, Makino, Sumita, Tamura, and
  Yokoo]{KawaseMaSu24}
Yasushi Kawase, Kazuhisa Makino, Hanna Sumita, Akihisa Tamura, and Makoto
  Yokoo.
\newblock Towards optimal subsidy bounds for envy-freeable allocations.
\newblock In \emph{Proceedings of the 38th AAAI Conference on Artificial
  Intelligence (AAAI)}, pages 9824--9831, 2024{\natexlab{a}}.

\bibitem[Kawase et~al.(2024{\natexlab{b}})Kawase, Nishimura, and
  Sumita]{KawaseNiSu24}
Yasushi Kawase, Koichi Nishimura, and Hanna Sumita.
\newblock Minimizing symmetric convex functions over hybrid of continuous and
  discrete convex sets.
\newblock In \emph{Proceedings of the 51st EATCS International Colloquium on
  Automata, Languages, and Programming (ICALP)}, pages 96:1--96:19,
  2024{\natexlab{b}}.

\bibitem[Klijn(2000)]{Klijn00}
Flip Klijn.
\newblock An algorithm for envy-free allocations in an economy with indivisible
  objects and money.
\newblock \emph{Social Choice and Welfare}, 17\penalty0 (2):\penalty0 201--215,
  2000.

\bibitem[Kulkarni et~al.(2021{\natexlab{a}})Kulkarni, Mehta, and
  Taki]{KulkarniMeTa21}
Rucha Kulkarni, Ruta Mehta, and Setareh Taki.
\newblock Indivisible mixed manna: On the computability of {MMS} + {PO}
  allocations.
\newblock In \emph{Proceedings of the 22nd ACM Conference on Economics and
  Computation (EC)}, pages 683--684, 2021{\natexlab{a}}.

\bibitem[Kulkarni et~al.(2021{\natexlab{b}})Kulkarni, Mehta, and
  Taki]{KulkarniMeTa21-AAAI}
Rucha Kulkarni, Ruta Mehta, and Setareh Taki.
\newblock On the {PTAS} for maximin shares in an indivisible mixed manna.
\newblock In \emph{Proceedings of the 35th AAAI Conference on Artificial
  Intelligence (AAAI)}, pages 5523--5530, 2021{\natexlab{b}}.

\bibitem[Kurokawa et~al.(2018)Kurokawa, Procaccia, and Wang]{KurokawaPrWa18}
David Kurokawa, Ariel~D. Procaccia, and Junxing Wang.
\newblock Fair enough: Guaranteeing approximate maximin shares.
\newblock \emph{Journal of the ACM}, 65\penalty0 (2):\penalty0 8:1--8:27, 2018.

\bibitem[Kyropoulou et~al.(2020)Kyropoulou, Suksompong, and
  Voudouris]{KyropoulouSuVo20}
Maria Kyropoulou, Warut Suksompong, and Alexandros~A. Voudouris.
\newblock Almost envy-freeness in group resource allocation.
\newblock \emph{Theoretical Computer Science}, 841:\penalty0 110--123, 2020.

\bibitem[Li et~al.(2022)Li, Li, and Wu]{LiLiWu22}
Bo~Li, Yingkai Li, and Xiaowei Wu.
\newblock Almost (weighted) proportional allocations for indivisible chores.
\newblock In \emph{Proceedings of the 31st ACM Web Conference (TheWebConf)},
  pages 122--131, 2022.

\bibitem[Li et~al.(2024{\natexlab{a}})Li, Li, Liu, and Wu]{LiLiLi24}
Bo~Li, Zihao Li, Shengxin Liu, and Zekai Wu.
\newblock Allocating mixed goods with customized fairness and indivisibility
  ratio.
\newblock In \emph{Proceedings of the 33rd International Joint Conference on
  Artificial Intelligence (IJCAI)}, pages 2868--2876, 2024{\natexlab{a}}.

\bibitem[Li et~al.(2015)Li, Zhang, and Zhang]{LiZhZh15}
Minming Li, Jialin Zhang, and Qiang Zhang.
\newblock Truthful cake cutting mechanisms with externalities: Do not make them
  care for others too much!
\newblock In \emph{Proceedings of the 24th International Conference on
  Artificial Intelligence (IJCAI)}, pages 589--595, 2015.

\bibitem[Li et~al.(2023)Li, Liu, Lu, and Tao]{LiLiLu23}
Zihao Li, Shengxin Liu, Xinhang Lu, and Biaoshuai Tao.
\newblock Truthful fair mechanisms for allocating mixed divisible and
  indivisible goods.
\newblock In \emph{Proceedings of the 32nd International Joint Conference on
  Artificial Intelligence (IJCAI)}, pages 2808--2816, 2023.

\bibitem[Li et~al.(2024{\natexlab{b}})Li, Liu, Lu, Tao, and Tao]{LiLiLu24}
Zihao Li, Shengxin Liu, Xinhang Lu, Biaoshuai Tao, and Yichen Tao.
\newblock A complete landscape for the price of envy-freeness.
\newblock In \emph{Proceedings of the 23rd International Conference on
  Autonomous Agents and Multiagent Systems (AAMAS)}, pages 1183--1191,
  2024{\natexlab{b}}.

\bibitem[Lindner and Rothe(2024)]{LindnerRo24}
Claudia Lindner and J\"{o}rg Rothe.
\newblock Cake-cutting: Fair division of divisible goods.
\newblock In J\"{o}rg Rothe, editor, \emph{Economics and Computation: An
  Introduction to Algorithmic Game Theory, Computational Social Choice, and
  Fair Division}, chapter~8, pages 507--603. Springer Cham, 2nd edition, 2024.

\bibitem[Lipton et~al.(2004)Lipton, Markakis, Mossel, and Saberi]{LiptonMaMo04}
Richard~J. Lipton, Evangelos Markakis, Elchanan Mossel, and Amin Saberi.
\newblock On approximately fair allocations of indivisible goods.
\newblock In \emph{Proceedings of the 5th ACM Conference on Electronic Commerce
  (EC)}, pages 125--131, 2004.

\bibitem[Liu et~al.(2024)Liu, Lu, Suzuki, and Walsh]{LiuLuSu24}
Shengxin Liu, Xinhang Lu, Mashbat Suzuki, and Toby Walsh.
\newblock Mixed fair division: {A} survey.
\newblock In \emph{Proceedings of the 38th AAAI Conference on Artificial
  Intelligence (AAAI)}, pages 22641--22649, 2024.
\newblock Senior Member Presentation Track.

\bibitem[Lu et~al.(2024)Lu, Peters, Aziz, Bei, and Suksompong]{LuPeAz24}
Xinhang Lu, Jannik Peters, Haris Aziz, Xiaohui Bei, and Warut Suksompong.
\newblock Approval-based voting with mixed goods.
\newblock \emph{Social Choice and Welfare}, 62\penalty0 (4):\penalty0 643--677,
  2024.

\bibitem[Maskin(1987)]{Maskin87}
Eric~S. Maskin.
\newblock On the fair allocation of indivisible goods.
\newblock In George~R. Feiwel, editor, \emph{Proceedings of the Arrow and the
  Foundations of the Theory of Economic Policy}, pages 341--349. Palgrave
  Macmillan UK, 1987.

\bibitem[Meunier and Zerbib(2019)]{MeunierZe19}
Fr\'{e}d\'{e}ric Meunier and Shira Zerbib.
\newblock Envy-free cake division without assuming the players prefer nonempty
  pieces.
\newblock \emph{Israel Journal of Mathematics}, 234\penalty0 (2):\penalty0
  907--925, 2019.

\bibitem[Moulin(2019)]{Moulin19}
Herv\'{e} Moulin.
\newblock Fair division in the internet age.
\newblock \emph{Annual Review of Economics}, 11\penalty0 (1):\penalty0
  407--441, 2019.

\bibitem[Narayan et~al.(2021)Narayan, Suzuki, and Vetta]{NarayanSuVe21}
Vishnu~V. Narayan, Mashbat Suzuki, and Adrian Vetta.
\newblock Two birds with one stone: Fairness and welfare via transfers.
\newblock In \emph{Proceedings of the 14th International Symposium on
  Algorithmic Game Theory (SAGT)}, pages 376--390, 2021.

\bibitem[Nguyen and Rothe(2023)]{NguyenRo23}
Trung~Thanh Nguyen and J\"{o}rg Rothe.
\newblock Complexity results and exact algorithms for fair division of
  indivisible items: {A} survey.
\newblock In \emph{Proceedings of the 32nd International Joint Conference on
  Artificial Intelligence (IJCAI)}, pages 6732--6740, 2023.
\newblock Survey Track.

\bibitem[Nishimura and Sumita(2023)]{NishimuraSu23}
Koichi Nishimura and Hanna Sumita.
\newblock Envy-freeness and maximum {N}ash welfare for mixed divisible and
  indivisible goods.
\newblock \emph{CoRR}, abs/2302.13342, 2023.

\bibitem[Plaut and Roughgarden(2020)]{PlautRo20}
Benjamin Plaut and Tim Roughgarden.
\newblock Almost envy-freeness with general valuations.
\newblock \emph{SIAM Journal on Discrete Mathematics}, 34\penalty0
  (2):\penalty0 1039--1068, 2020.

\bibitem[Procaccia(2016)]{Procaccia16}
Ariel~D. Procaccia.
\newblock Cake cutting algorithms.
\newblock In Felix Brandt, Vincent Conitzer, Ulle Endriss, J\'{e}r\^{o}me Lang,
  and Ariel~D. Procaccia, editors, \emph{Handbook of Computational Social
  Choice}, chapter~13, pages 311--329. Cambridge University Press, 2016.

\bibitem[Procaccia(2020)]{Procaccia20}
Ariel~D. Procaccia.
\newblock An answer to fair division's most enigmatic question: Technical
  perspective.
\newblock \emph{Communications of the ACM}, 63\penalty0 (4):\penalty0 118,
  2020.

\bibitem[Rey and Maly(2023)]{ReyMa23}
Simon Rey and Jan Maly.
\newblock The (computational) social choice take on indivisible participatory
  budgeting.
\newblock \emph{CoRR}, abs/2303.00621, 2023.

\bibitem[Robertson and Webb(1998)]{RobertsonWe98}
Jack Robertson and William Webb.
\newblock \emph{Cake-Cutting Algorithm: Be Fair If You Can}.
\newblock A K Peters/CRC Press, 1998.

\bibitem[Rothe(2024)]{Rothe24}
J\"{o}rg Rothe, editor.
\newblock \emph{Economics and Computation: An Introduction to Algorithmic Game
  Theory, Computational Social Choice, and Fair Division}.
\newblock Springer Cham, 2nd edition, 2024.

\bibitem[Seddighin et~al.(2021)Seddighin, Saleh, and Ghodsi]{SeddighinSaGh21}
Masoud Seddighin, Hamed Saleh, and Mohammad Ghodsi.
\newblock Maximin share guarantee for goods with positive externalities.
\newblock \emph{Social Choice and Welfare}, 56\penalty0 (2):\penalty0 291--324,
  2021.

\bibitem[Segal-Halevi(2018)]{Segal-Halevi18}
Erel Segal-Halevi.
\newblock Fairly dividing a cake after some parts were burnt in the oven.
\newblock In \emph{Proceedings of the 17th International Conference on
  Autonomous Agents and Multiagent Systems (AAMAS)}, pages 1276--1284, 2018.

\bibitem[Shah(2017)]{Shah17}
Nisarg Shah.
\newblock Making the world fairer.
\newblock \emph{XRDS: Crossroads, The ACM Magazine for Students}, 24\penalty0
  (1):\penalty0 24--28, 2017.

\bibitem[Skowron and G\'{o}recki(2022)]{SkowronGo22}
Piotr Skowron and Adrian G\'{o}recki.
\newblock Proportional public decisions.
\newblock In \emph{Proceedings of the 36th AAAI Conference on Artificial
  Intelligence (AAAI)}, pages 5191--5198, 2022.

\bibitem[Steinhaus(1948)]{Steinhaus48}
Hugo Steinhaus.
\newblock The problem of fair division.
\newblock \emph{Econometrica}, 16\penalty0 (1):\penalty0 101--104, 1948.

\bibitem[Su(1999)]{Su99}
Francis~Edward Su.
\newblock Rental harmony: {S}perner's lemma in fair division.
\newblock \emph{The American Mathematical Monthly}, 106\penalty0 (10):\penalty0
  930--942, 1999.

\bibitem[Suksompong(2021)]{Suksompong21}
Warut Suksompong.
\newblock Constraints in fair division.
\newblock \emph{ACM SIGecom Exchanges}, 19\penalty0 (2):\penalty0 46--61, 2021.

\bibitem[Suksompong(2025)]{Suksompong25}
Warut Suksompong.
\newblock Weighted fair division of indivisible items: {A} review.
\newblock \emph{Information Processing Letters}, 187:\penalty0 106519, 2025.

\bibitem[Suksompong and Teh(2022)]{SuksompongTe22}
Warut Suksompong and Nicholas Teh.
\newblock On maximum weighted {N}ash welfare for binary valuations.
\newblock \emph{Mathematical Social Sciences}, 117:\penalty0 101--108, 2022.

\bibitem[Suksompong and Teh(2023)]{SuksompongTe23}
Warut Suksompong and Nicholas Teh.
\newblock Weighted fair division with matroid-rank valuations: Monotonicity and
  strategyproofness.
\newblock \emph{Mathematical Social Sciences}, 126:\penalty0 48--59, 2023.

\bibitem[Sun et~al.(2023)Sun, Chen, and Doan]{SunChDo23}
Ankang Sun, Bo~Chen, and Xuan~Vinh Doan.
\newblock Equitability and welfare maximization for allocating indivisible
  items.
\newblock \emph{Autonomous Agents and Multi-Agent Systems}, 37\penalty0
  (1):\penalty0 8:1--8:45, 2023.

\bibitem[Suzuki and Vollen(2024)]{SuzukiVo24}
Mashbat Suzuki and Jeremy Vollen.
\newblock Maximum flow is fair: {A} network flow approach to committee voting.
\newblock In \emph{Proceedings of the 25th ACM Conference on Economics and
  Computation (EC)}, 2024.
\newblock Forthcoming.

\bibitem[Tinbergen(1930)]{Tinbergen30}
Jan Tinbergen.
\newblock Mathematiese psychologie.
\newblock \emph{Mens en Maatschappij}, 6\penalty0 (4):\penalty0 342--352, 1930.

\bibitem[Varian(1974)]{Varian74}
Hal~R. Varian.
\newblock Equity, envy, and efficiency.
\newblock \emph{Journal of Economic Theory}, 9\penalty0 (1):\penalty0 63--91,
  1974.

\bibitem[Viswanathan and Zick(2023)]{ViswanathanZi23}
Vignesh Viswanathan and Yair Zick.
\newblock A general framework for fair allocation under matroid rank
  valuations.
\newblock In \emph{Proceedings of the 24th ACM Conference on Economics and
  Computation (EC)}, pages 1129--1152, 2023.

\bibitem[Walsh(2020)]{Walsh20}
Toby Walsh.
\newblock Fair division: {T}he computer scientist's perspective.
\newblock In \emph{Proceedings of the 29th International Joint Conference on
  Artificial Intelligence (IJCAI)}, pages 4966--4972, 2020.

\bibitem[Woeginger(1997)]{Woeginger97}
Gerhard~J. Woeginger.
\newblock A polynomial-time approximation scheme for maximizing the minimum
  machine completion time.
\newblock \emph{Operations Research Letters}, 20\penalty0 (4):\penalty0
  149--154, 1997.

\bibitem[Wu and Zhou(2024)]{WuZh24}
Xiaowei Wu and Shengwei Zhou.
\newblock Tree splitting based rounding scheme for weighted proportional
  allocations with subsidy.
\newblock \emph{CoRR}, abs/2404.07707, 2024.

\bibitem[Wu et~al.(2023)Wu, Zhang, and Zhou]{WuZhZh23}
Xiaowei Wu, Cong Zhang, and Shengwei Zhou.
\newblock One quarter each (on average) ensures proportionality.
\newblock In \emph{Proceedings of the 19th Conference on Web and Internet
  Economics (WINE)}, pages 582--599, 2023.

\bibitem[Zhou and Wu(2024)]{ZhouWu24}
Shengwei Zhou and Xiaowei Wu.
\newblock Approximately {EFX} allocations for indivisible chores.
\newblock \emph{Artificial Intelligence}, 326:\penalty0 104037, 2024.

\end{thebibliography}
\end{document}